\documentclass[
 reprint,twocolumn,
 groupedaddress,
 superscriptaddress,
 aps,floatfix
]{revtex4-2}
\usepackage{graphicx}
\usepackage{dcolumn}
\usepackage{bm}

\usepackage{amssymb,amsmath,amsfonts,latexsym,graphicx,verbatim}

\usepackage[english]{babel}
\usepackage{times}
\usepackage{latexsym}
\usepackage{fancyhdr}
\usepackage{float}
\usepackage{afterpage}
\usepackage{listings}
\usepackage{multirow}
\usepackage[table]{xcolor}

\usepackage{bbm}
\usepackage{upgreek}

\definecolor{Ablue}{rgb}{0.96,0.24,0.00}

\definecolor{Abluetitle}{rgb}{0.,0.24,0.51}
\newcommand{\bluetitle}{\color{Abluetitle}}

\begin{document}

\title{\bluetitle{Large Room Temperature Bulk DNP of $^{13}$C via P1 Centers in Diamond}}

\author{Daphna Shimon}
\email{daphna.shimon@mail.huji.ac.il}
\affiliation{Institute of Chemistry, The Hebrew University of Jerusalem, Edmond J. Safra, Givat Ram, Jerusalem, Israel.}
\author{Kelly A. Cantwell}
\affiliation{Department of Physics and Astronomy, Dartmouth College, Hanover, NH 03755, U.S.A.}
\author{Linta Joseph}
\affiliation{Department of Physics and Astronomy, Dartmouth College, Hanover, NH 03755, U.S.A.}
\author{Ethan Q. Williams}
\affiliation{Department of Physics and Astronomy, Dartmouth College, Hanover, NH 03755, U.S.A.}
\author{Zaili Peng}
\affiliation{Department of Chemistry, University of Southern California, Los Angeles, California 90089, U.S.A.}
\author{Susumu Takahashi}
\affiliation{Department of Chemistry, University of Southern California, Los Angeles, California 90089, U.S.A.}
\affiliation{Department of Physics and Astronomy, University of Southern California, Los Angeles, California 90089, USA}
\author{Chandrasekhar Ramanathan}
\email{chandrasekhar.ramanathan@dartmouth.edu}
\affiliation{Department of Physics and Astronomy, Dartmouth College, Hanover, NH 03755, U.S.A.}

\begin{abstract}
   We use microwave-induced dynamic nuclear polarization (DNP) of the substitutional nitrogen defects (P1 centers) in diamond to hyperpolarize bulk $^{13}$C nuclei in both single crystal and powder samples at room temperature at 3.34 T. The  large ($>100$-fold) enhancements demonstrated correspond to a greater than 10,000 fold improvement in terms of signal averaging of the 1\% abundant $^{13}$C spins. The DNP was performed using low-power solid state sources under static (non-spinning) conditions. The DNP spectrum (DNP enhancement as a function of microwave frequency) of diamond powder shows features that broadly correlate with the EPR spectrum.  A well-defined negative Overhauser peak and two solid effect peaks are observed for the central ($m_I=0$) manifold of the $^{14}$N spins. Previous low temperature measurements in diamond had measured a positive Overhauser enhancement in this manifold. Frequency-chirped millimeter-wave excitation of the electron spins is seen to significantly improve the enhancements for the two outer nuclear spin manifolds ($m_I = \pm 1$) and to blur some of the sharper features associated with the central manifolds. The outer lines are best fit using a combination of the cross effect and a truncated cross effect -- which is known to mimic features of an Overhauser effect. Similar features are also observed in experiments on single crystal samples.  The observation of all of these mechanisms in a single material system under the same experimental conditions is likely due to the significant heterogeneity of the high pressure, high temperature (HPHT) type Ib diamond samples used. Large room temperature DNP enhancements at fields above a few Tesla enable spectroscopic studies with better chemical shift resolution under ambient conditions.
\end{abstract}

\maketitle


\section{Introduction} \label{sec:intro}
\noindent Nuclear magnetic resonance (NMR) spectroscopy shows exquisite chemical sensitivity when reporting on the local magnetic environments of the spins at atomic scales.  However, it's low detection sensitivity has long required the use of relatively large sample volumes.  Microwave-induced dynamic nuclear polarization (DNP), a technique to produce large nuclear spin signal enhancements via polarization transfer from electrons \cite{Hovav2010, Hovav2012_CE, Shimon2012, Thankamony2017}, is driving a technological revolution by enabling NMR and magnetic resonance imaging (MRI) studies of low abundance, low-gamma spins and nuclei at surfaces and interfaces for the first time \cite{corzilius2020high,rossini2013dynamic}.

The electron-nuclear polarization transfer step in most DNP experiments is typically performed at cryogenic temperatures in order to reduce the electron spin-lattice relaxation rates below the strength of the hyperfine interactions.   This is true for both dissolution DNP, which 
is increasingly being used to study room-temperature phenomena in biomedical systems \cite{zhang2018applications}, as well as for high-field magic-angle sample spinning (MAS) DNP which has enabled high-sensitivity, high-resolution NMR studies of chemistry in sample-limited solid systems \cite{corzilius2020high}.  Large room temperature DNP enhancements at high field would enable high-resolution NMR spectroscopic studies under ambient conditions.   

The nitrogen-vacancy (NV) center in diamond is a promising electron spin system for room temperature DNP applications at low field because of its long coherence and relaxation times  \cite{Doherty-2013,Schirhagl-2014,Degen-2017,Suter-2017,Smits-2019}. The electron spin of the NV center can be optically polarized to near unity polarization at room temperature.  NV centers can also be detected (even at the single spin level) and coherently manipulated using optically-detected magnetic resonance (ODMR).   In addition to hyperpolarizing the $^{13}$C spins locally around a single NV center \cite{Jacques2009,Smeltzer2009,Childress2013}, the hyperpolarization of bulk samples via ensembles of NV centers has been demonstrated, both at low field \cite{Fischer-2013,Ajoy-2018SciAdv,Ajoy2018} and at high magnetic fields \cite{King-2010,Drake-2015,Scott-2016}.   However, at high magnetic fields, the magnitude and sign of the $^{13}$C hyperpolarization were seen to depend strongly on the orientation of the diamond crystal \cite{Scott-2016} making it difficult to hyperpolarize bulk powders. 
Nanodiamonds are chemically stable and their surfaces can be chemically functionalized, making them excellent candidates  as local NMR probes \cite{Rej2015,Chen-2017} -- similar to silicon particles  \cite{dementyev2008dynamic,aptekar2009silicon,cassidy2013vivo,cassidy2013radical} -- as well as fluorescent biomarkers \cite{Schirhagl-2014}.  A key challenge to the practical use of DNP via defects in diamond is the need to achieve high polarizations outside the diamond surface.  While the hyperpolarization of spins on the surface of diamond samples has been demonstrated at low magnetic fields \cite{Rej2016,Shagieva2018,Broadway-2018, Acebal-2018}, the transfer of polarization to external spins remains challenging.
Relaxation due to surface defects and slow spin diffusion in natural abundance $^{13}$C samples have so far been the key limiting factors to achieving high polarizations.  


In this work we study microwave-induced DNP of diamond at room temperature at 3.34 T under static (non-spinning) conditions.  The P1 center is a spin-1/2 substitutional nitrogen impurity in the diamond lattice that also exhibits long coherence and relaxation times \cite{Reynhardt2001, Cox1994, Takahashi-2008}, though it is not optically active.  There are typically at least an order of magnitude more P1 centers in a diamond crystal than NV centers \cite{Takahashi-2008}, though the best efficiencies for converting P1 to NV centers can reach up to 20-25\% \cite{Mindarava-2020,Kollarics-2022}. As a consequence, most $^{13}$C nuclear spins in diamond are much closer to a P1 center than an NV center, making the P1s potentially more efficient polarization sources. 
At high magnetic fields, DNP via P1 centers can produce a significant increase in nuclear spin polarization at cryogenic temperatures \cite{Lock1992, Reynhardt1998,Reynhardt2001,Casabianca2011,Bretschneider2016,Kwiatkowski2018}. Bretschneider et al.\ have also reported a room temperature $^{13}$C DNP enhancement of 130 at 9.4 T under 8 kHz MAS conditions with 10 W of microwave power at 263 GHz \cite{Bretschneider2016}. Other high-field room temperature enhancements have also been reported \cite{Carroll-2018}.
Hyperpolarization via P1 centers is also well suited to applications when optical access is not possible.   Here, we show a greater than 100-fold enhancement of the $^{13}$C spins in both single crystal and powder diamond samples.  In contrast to earlier work, these enhancements were achieved using a solid-state source with about 240 mW of power.  The DNP spectrum exhibits multiple features indicating that several different DNP mechanisms are operational in these systems. We elucidate the different physical mechanisms and explain their role in both single crystal and powder samples of diamond.   Understanding these mechanisms could guide the design of more effective hyperpolarization strategies and the improved design of diamond substrates for polarizing external spins.


\begin{figure}
\includegraphics[width=0.49\textwidth]{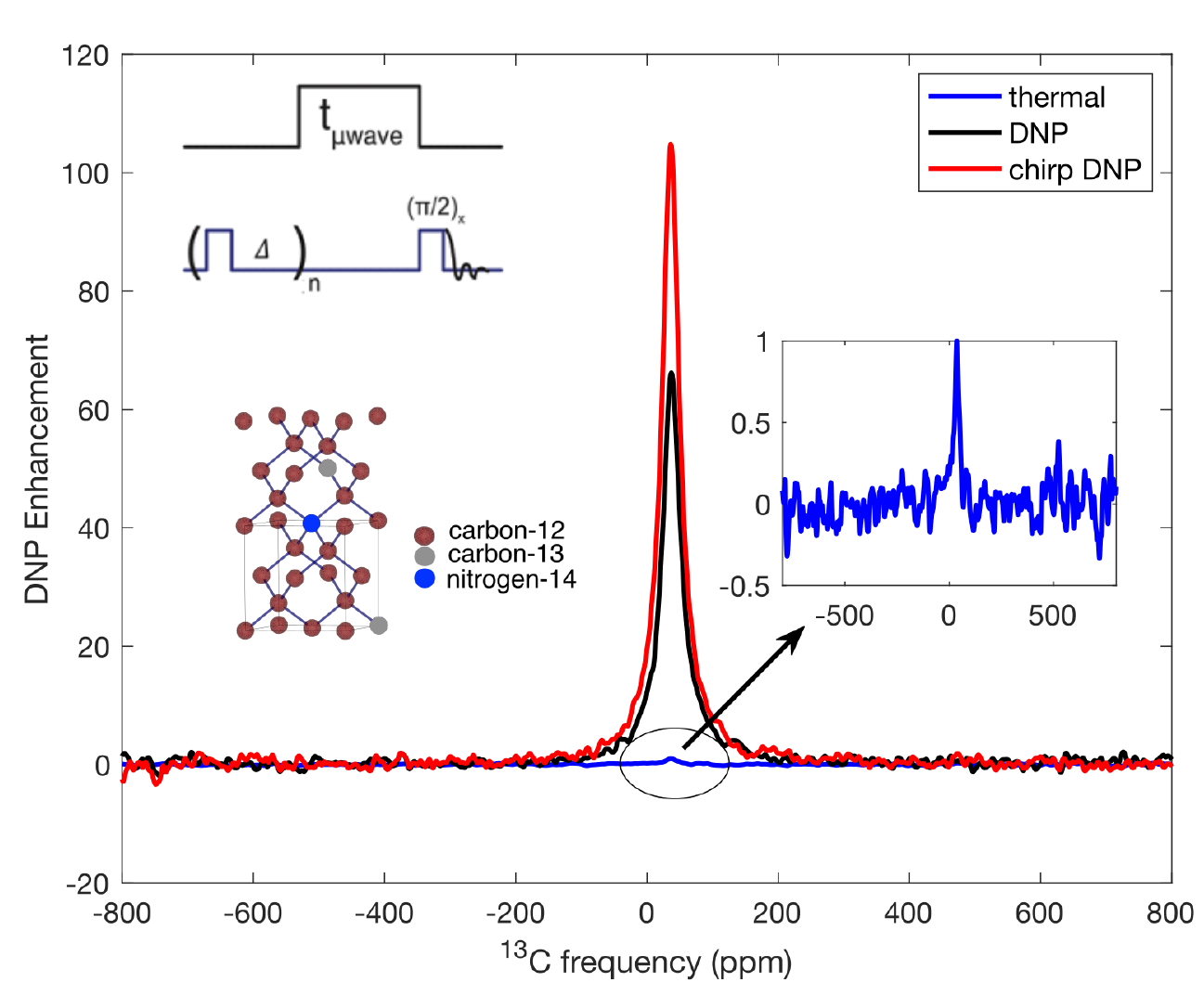}
\caption{The $^{13}$C NMR spectrum of the diamond powder, with MW irradiation (black line) at 93.696 GHz and without MW irradiation (blue line), resulting in DNP enhancement of 75. With the chirped excitation the enhancement is seen to rise to 114 (red line). The signal was recorded after a recycle delay of 3000 s for the thermal signal, and after 3000 s of MW irradiation for the DNP and chirp DNP signals. Triangular ramp-up modulation was used with a 5 kHz modulation frequency and a 117 MHz modulation amplitude. The pulse sequence used to record the NMR spectra is shown in the inset, with $\Delta$ being the delay in the saturation pulse train and $t_{\mu\text{wave}}$ the DNP build-up time.  The model of the diamond lattice shown in the figure was created using VESTA \cite{Vesta}.}
 \label{fig:NMR_spectrum}
\end{figure}

The paper is organized as follows: Our main experimental results on the powder sample are described  in Section~\ref{sec:main_dnp}, followed by a brief overview of relevant DNP mechanisms in Section~\ref{sec:mechanisms}. We discuss the fit of these mechanisms to the observed DNP spectra in both the single crystal and powder samples in Section~\ref{sec:fitting_mechanisms}, discuss how sample heterogeneity gives rise to the different mechanisms in Section~\ref{sec:heterogeneity}, and conclude in Section~\ref{sec:conclusions}.

\begin{figure}
\includegraphics[width=0.49\textwidth]{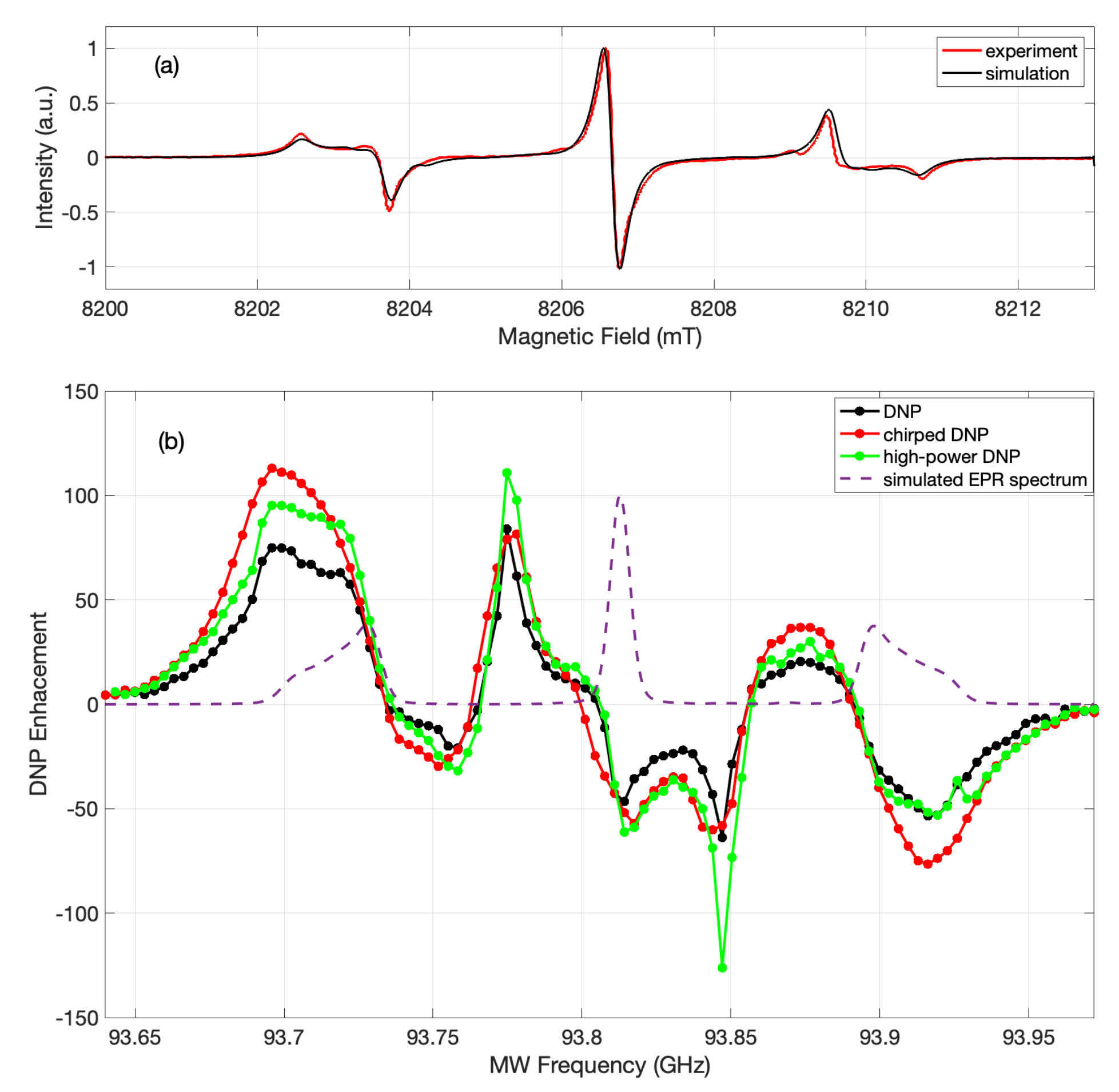}
\caption{
(a) The high-field EPR spectrum of the diamond powder, and the best fit according to EasySpin. 
The simulation of the experimental EPR line used a 0.19 mT Lorentzian line broadening.
(b) The $^{13}$C DNP spectrum of the diamond powder: enhancement measured as a function of constant frequency MW (black symbols) and chirped frequency MW irradiation (red symbols) with 240 mW power; and constant frequency MW at 500 mw power (green symbols). The figure also shows simulated EPR spectrum under the experimental conditions (dashed magenta line). The simulation of the expected EPR line at 94 GHz used a 6 MHz line broadening.  Error bars are not shown (the $\sim$ 10 \% errors in enhancement are dominated by the SNR of the thermal signal). }
 \label{fig:DNP_spectrum}
\end{figure}

\section{Using P1 Centers for DNP at Room Temperature} \label{sec:main_dnp}
\noindent Our experiments were performed at room temperature at 3.34 T, where the $^{13}$C Larmor frequency is 35.8 MHz.  Figure~\ref{fig:NMR_spectrum} shows the thermal equilibrium $^{13}$C signal from a bulk diamond powder sample (Element 6) after 128 averages with a recycle delay of 3000 s. The  The signal was acquired using the pulse sequence shown in the inset of Figure~\ref{fig:NMR_spectrum}.   The type Ib diamond is made by high pressure, high temperature (HPHT) synthesis. The diamond microparticles are 15--25 $\mu$m in diameter and are specified to have a nitrogen concentration less than 200 ppm (Element 6 estimated the actual values were about 110-130 ppm).

The figure also shows the single shot DNP signal with the same 3000 s build-up time under constant frequency irradiation and chirped millimeter-wave (MW) irradiation centered around the frequency of 93.696 GHz. At this MW excitation frequency the constant frequency DNP excitation resulted in a DNP enhancement of $75\pm8$ and the chirped excitation results in a DNP enhancement of $114\pm11$. The chirped excitation used a triangular ramp-up function with a 5 kHz modulation frequency and a 117 MHz modulation amplitude. The maximum output power of our MW source is about 240 mW (see Materials and Methods - Section VI.C).  

All NMR spectra show a single peak at 37$\pm$1 ppm, which matches the literature value for $^{13}$C nuclei in diamonds \cite{Reynhardt2001} (see Figure~\ref{fig:NMR_spectrum}). The NMR peak has a width of 1.12 kHz showing significant inhomogeneous line-broadening. It is possible to detect several hundred echoes in a pulsed spin-lock experiment as described in the Section~\ref{SISec:decoherence} of the Supplementary  Information (SI).   Stroboscopic detection of multiple echoes would
significantly improve the signal-to-noise ration (SNR).

Figure \ref{fig:DNP_spectrum}(a) shows the experimentally measured continuous-wave (CW) EPR spectrum of the sample measured at room temperature using the 230 GHz EPR system at the University of Southern California \cite{Cho-2014,Cho-2015}. The EPR spectrum features three lines, consisting of a single electron split by the hyperfine coupling to the spin-1 $^{14}$N nucleus of the P1 center.
The anisotropic hyperfine interaction results in a powder broadening of the $m_I = \pm 1$ manifolds.  The figure also shows an EasySpin \cite{stoll2006} quantum mechanical simulation of the spectrum for this sample  overlaid on the experimental spectrum. Each P1 center was modeled as an e-$^{14}$N system, with an isotropic g-factor, g=2.0024, hyperfine coupling strengths with the $^{14}$N nucleus with principal axis components $A_{x}^N = A_{y}^N = 82$ MHz, $A_{z}^N$ = 114 MHz \cite{Loubser-1978} (see Section~\ref{SISec:simulateEPR} of the SI).
The $^{14}$N nuclear quadrupolar interaction was neglected.

Figure \ref{fig:DNP_spectrum}(b) shows the DNP spectrum (DNP enhancement as a function of millimeter wave frequency) for $^{13}$C obtained with a build-up time of 3000 s under three  different experimental conditions. These include (i) constant frequency MW excitation with $\sim 240$ mW power; (ii) frequency-chirped  MW excitation with $\sim 240$ mW power using the same triangular ramp-up function described above; and (iii) constant frequency MW excitation with $\sim 500$ mW power. The higher power sweep was obtained by using the attenuated output of the 240 mW source to injection-lock an impatt diode source.  Additional details are provided in the Supplementary Information. The figure also shows the simulated EPR spectrum at 3.34 T using the same sample parameters used to fit the high-field spectra -- see Section~\ref{SISec:simulateEPR} of the SI for details.  {The estimated uncertainty in the enhancement is about 10\%, which is dominated by the standard deviation of the thermal signal.}

\section{DNP Mechanisms} \label{sec:mechanisms}
\begin{figure}[t]
\includegraphics[width=0.48\textwidth]{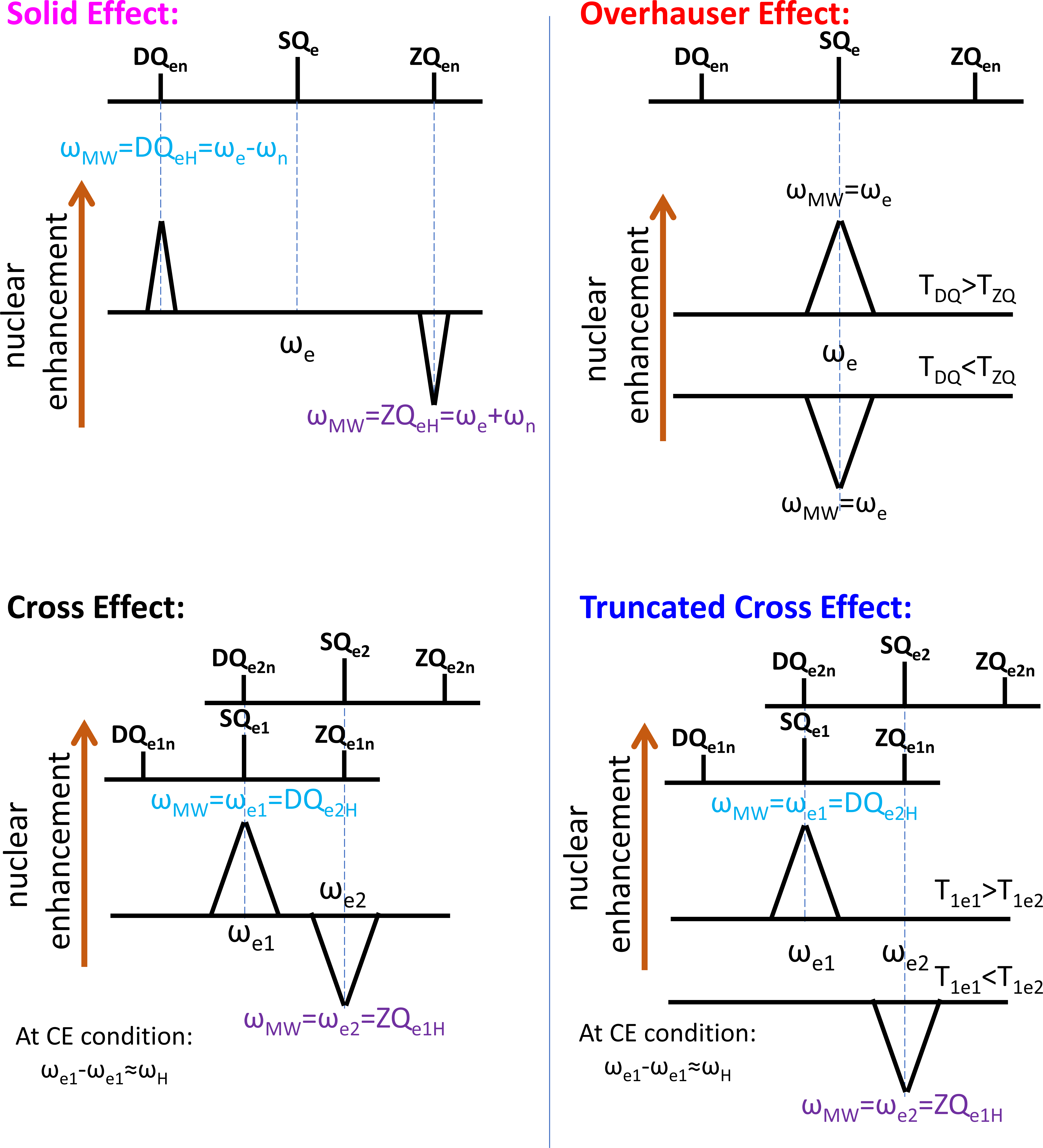}
\caption{Schematic DNP enhancement spectra showing the positions of the SE-DNP (e-n system), OE-DNP (e-n system), CE-DNP (e-e-n system) and truncated CE-DNP (e-e-n system) mechanisms with respect to the EPR transitions in each system. The DQ and ZQ transition labels represent a nucleus with a positive gyromagnetic ratio, and an electron with a negative gyromagnetic ratio.
\label{fig:DNP_mechanisms}}
\end{figure}

\noindent While the DNP spectrum in Figure \ref{fig:DNP_spectrum}(b) is seen to broadly correlate with the EPR spectrum, it is not immediately possible to identify the underlying DNP mechanism(s).  Here we outline the different DNP mechanisms that could play a role. Most solid systems studied by DNP typically exhibit hyperpolarization via one of five mechanisms \cite{abragam1961principles,kundu2019dnp,Hovav2010,Hovav2012_CE}. 
Figure \ref{fig:DNP_mechanisms} schematically illustrates the first four mechanisms. In broad EPR lines, positive and negative enhancement from the same mechanism and different mechanisms often overlap \cite{Kaminker2017,Kaminker2018,Shimon2012}. Additional smaller DNP features have also been observed due to the higher-order multi-spin processes \cite{shimon2015simultaneous}.

\vspace*{-0.2in}
\subsubsection{Overhauser Effect (OE)}
\vspace*{-0.2in}
The OE is typically observed in metals and liquids, systems in which the hyperfine interactions are strongly modulated in time. It has been shown, however, that the OE can also be observed in dielectric systems with strong localized exchange interactions \cite{Dementyev-2011}, and in mixed-valence organic radicals such as BDPA \cite{Can2014, Pylaeva2017, Ji2018, pylaeva2021}. Fluctuations of the hyperfine interaction result in electron-nuclear (e-n) cross-relaxation at the double quantum (DQ: $\omega_e - \omega_n$) or zero quantum (ZQ: $\omega_e + \omega_n$) e-n transitions, with characteristic decay times T$_{1DQ}$ and T$_{1ZQ}$, respectively \cite{Can2014, Pylaeva2017, Ji2018}.  When irradiating directly on the electron single quantum (SQ) transitions, an imbalance between T$_{1DQ}$ and T$_{1ZQ}$ results in nuclear enhancement, with positive enhancement when T$_{1DQ}$ $>$ T$_{1ZQ}$ and negative enhancement when T$_{1DQ}$ $<$ T$_{1ZQ}$.

\vspace*{-0.2in}
\subsubsection{Solid Effect (SE)}
\vspace*{-0.2in}
The SE is typically observed in isolated electron-nuclear spin systems in insulators where anisotropic hyperfine interactions admix the nuclear spin states. MW irradiation of the nominally  ``forbidden'' DQ (positive enhancement) and ZQ (negative enhancement) transitions leads to the enhancement of the nuclear spin polarization \cite{abragam1961principles,kundu2019dnp,Hovav2010}. If the ESR line is inhomogeneously broadened, frequency or field modulation can produce an integrated solid effect where the enhancements of the DQ and ZQ become additive under the appropriate modulation conditions \cite{Henstra-1988}.

\vspace*{-0.2in}
\subsubsection{Cross Effect (CE)}
\vspace*{-0.2in}
At higher electron spin concentrations, the three-spin (two electrons + one nucleus) CE process results from microwave irradiation of an inhomogeneously broadened EPR line \cite{abragam1961principles,kundu2019dnp}. The CE-DNP mechanism results in nuclear hyperpolarization when two electrons fulfill the so called CE condition ($\omega_{e1}-\omega_{e2}=\omega_{n}$ for $\omega_{e1}>\omega_{e2}$) and have unequal polarizations (usually due to irradiation of one of the electrons) \cite{Hovav2012_CE, Hovav2015}. 

\vspace*{-0.2in}
\subsubsection{Truncated Cross Effect (tCE)}
\vspace*{-0.2in}
Recently, Equbal {\em et al}.\ observed that it is possible for the CE to masquerade as an OE when the CE condition is satisfied by two pools of electrons, one with a very fast T$_{1e}$ relaxation and the other with much slower T$_{1e}$ relaxation \cite{equbal2018truncated}. When this happens, irradiating on the electrons that exhibit slow relaxation results in nuclear enhancement, due to the saturation of these electrons, and the formation of a polarization difference between the two pools of electrons. However, irradiating on the electrons that exhibit fast relaxation does not result in nuclear enhancement, because the electrons in the fast relaxing pool cannot be saturated. As a result, only positive or negative enhancement will be observed (depending on the MW frequencies of the fast and slow electrons), and the CE will appear truncated. For this truncated CE, the DNP enhancement is directly observed at the EPR frequency of the electron pool with the slow relaxation. 

\vspace*{-0.2in}
\subsubsection{Thermal Mixing (TM)}
\vspace*{-0.2in}
Finally, TM is a statistical thermodynamics description for many coupled electron spins, and requires MW irradiation directly on a homogeneously broadened EPR line \cite{abragam1961principles,kundu2019dnp}.  Note that TM is not expected to be significant at the concentration of P1 centers present in this sample.


\section{Analysis of the DNP Spectra}  \label{sec:fitting_mechanisms}

\subsection{Single Crystal}
\noindent In order to better understand the underlying mechanisms we first attempted to fit the DNP spectrum obtained from a single HPHT Type Ib sample (Element 6) specified with a nominal nitrogen concentration $< 200$ ppm.  Fitting the single crystal data should be simpler as we do not observe the effects of powder averaging of the anisotropic nitrogen hyperfine interactions.  We were unable to measure a thermal NMR signal in the single crystal sample even with extensive averaging.  We estimate a lower bound to the maximum enhancement of about 180 determined by the signal-to-noise of the DNP signal (see Section~\ref{SISec:singleXtal} of the SI).

\begin{figure}[htb]
\includegraphics[width=0.48\textwidth]{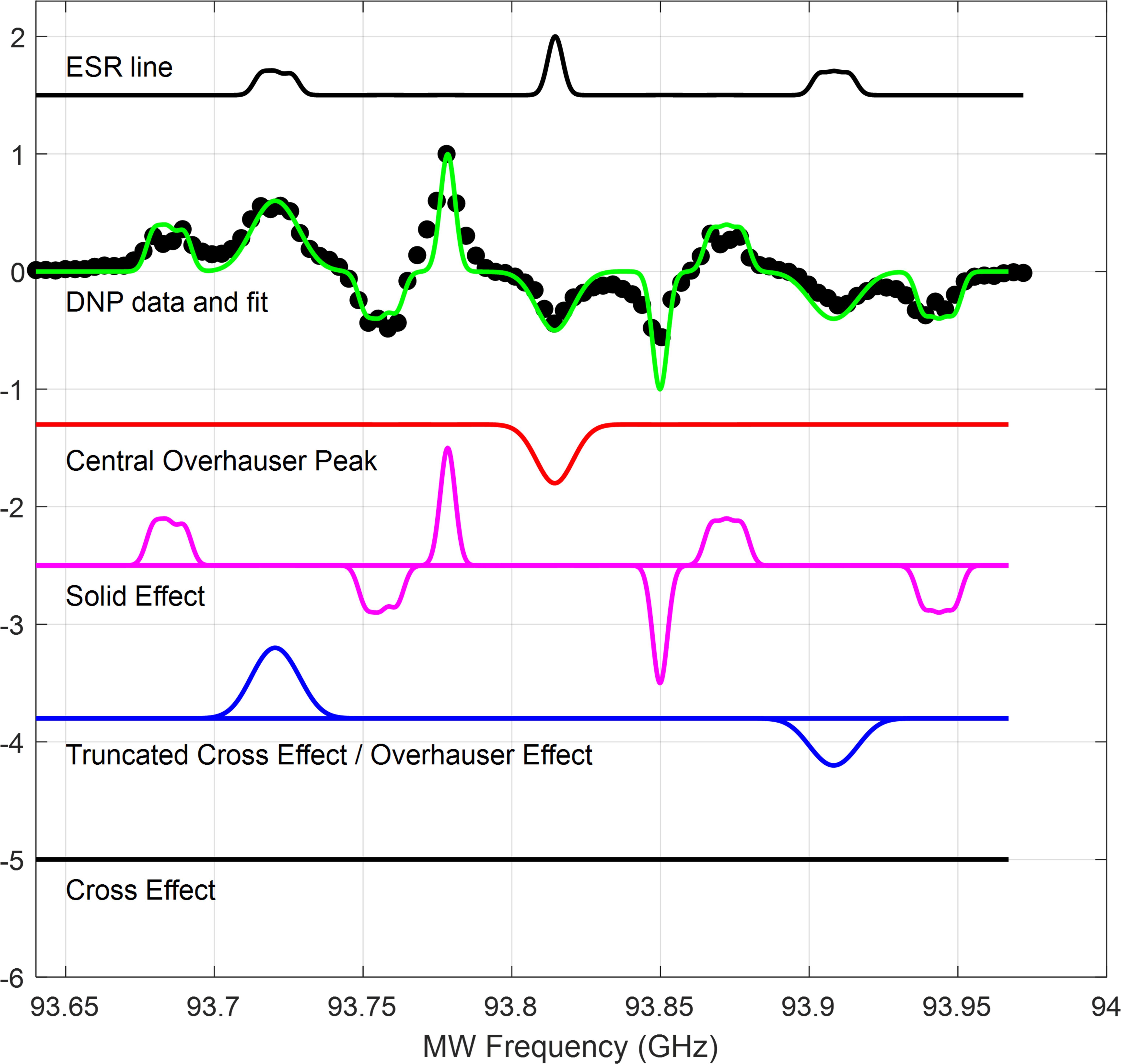}
\caption{Fit of experimental constant-frequency DNP spectrum (black circles) for a single-crystal diamond sample.  The fit uses a sum (green line) of the OE (red line), SE (magenta line), the truncated CE (blue line), and the CE (black line). The EPR line plotted above includes line broadening of 10 MHz of the outer EPR lines. The single crystal EPR line was simulated with EasySpin using the Euler angles
$\alpha=54.0013^{\circ}$, $\beta=134.4274^{\circ}$, $\gamma=18.0023^{\circ}$.
\label{fig:DNP_spectrum_xtal_fit}}
\end{figure}

Fig.\ \ref{fig:DNP_spectrum_xtal_fit} shows the DNP spectrum obtained from the single crystal sample.   The spectrum shows well-resolved peaks in the $m_I=\pm1$ manifold of the nitrogen spins which allows us to easily simulate the expected ESR spectrum at this crystal orientation.  The additional electron of the P1 center lies along one of the four equivalent C-N bonds due to the Jahn-Teller distortion. The EPR spectrum of a single diamond crystal can show between 3 and 7 distinct peaks, depending on the orientation of the crystal with respect to the external field. 

In Fig.\ \ref{fig:DNP_spectrum_xtal_fit}, we have fit the DNP spectrum using a very crude method of convolving the EPR line with delta functions to form the basic shapes for the SE, CE and tCE/OE DNP mechanisms, and adjusting their amplitudes to achieve the best agreement with the experimental spectrum \cite{Banerjee2013}. The shapes were constructed for each EPR line separately, and then the relative amplitudes were adjusted. For a detailed description of this method see Section~\ref{SISec:FittingDNP} of the SI.   


In the central nuclear spin manifold ($m_I = 0$), the sharp features correspond to an OE peak at 93.81 GHz and SE peaks (at the $^{13}$C nuclear side-band frequencies) at 93.78 and 93.85 GHz \cite{abragam1961principles,kundu2019dnp}. The origin of the temporal modulation that drives OE enhancement is still not clear. One potential source is the dynamic Jahn-Teller distortion. Early ENDOR studies suggested a reorientation rate less than 3.5 GHz at room temperature \cite{cook1966electron}. Ammerlaan and Bergmeister studied the reorientation rate of the P1 defect due to thermal excitation and tunneling in the temperature regime between 78 K and 200 K \cite{Ammerlaan-1981} while Loubser and van Ryneveld studied the rate between 600 and 1230 K \cite{loubser1967dynamic}. Interpolating between these sets of experimental results suggests a  characteristic reorientation rate on the order of 1--10 Hz at 300 K. This is likely to be too low to drive the observed Overhauser effect as it is necessary to have temporal modulations at the $^{13}$C Larmor frequency to induce an OE. Another possibility is that the OE is induced by exchange or dipolar-coupled clusters of P1 centers \cite{Abragam-1958,Dementyev-2011}.

The outer lines are best fit using the OE or truncated CE. If it is in fact the OE, the experimental spectrum necessitates that the OE on the low frequency side be positive, while the OE on the high frequency side be negative. To our knowledge, this type of effect has never been reported previously, and the origin of the difference in sign is unclear. Therefore, we also suggest that we may be in fact observing two truncated CE peaks.

The truncated CE is possible in this system if we have an additional pool of very fast relaxing P1 centers that appear in the frequency range between the outer manifolds (93.76 GHz to 93.89 GHz). If this were the case, then irradiating on the low frequency EPR line should result in positive enhancement (because we would be irradiating on the lower frequency EPR line of the CE pair), and irradiating on the high frequency EPR line should result in negative enhancement (because we would be irradiating on the higher frequency EPR line of the CE pair), exactly as observed. These fast relaxing spins would be difficult to observe in a standard CW-EPR experiment.  

\begin{figure}[ht]
\includegraphics[width=0.48\textwidth]{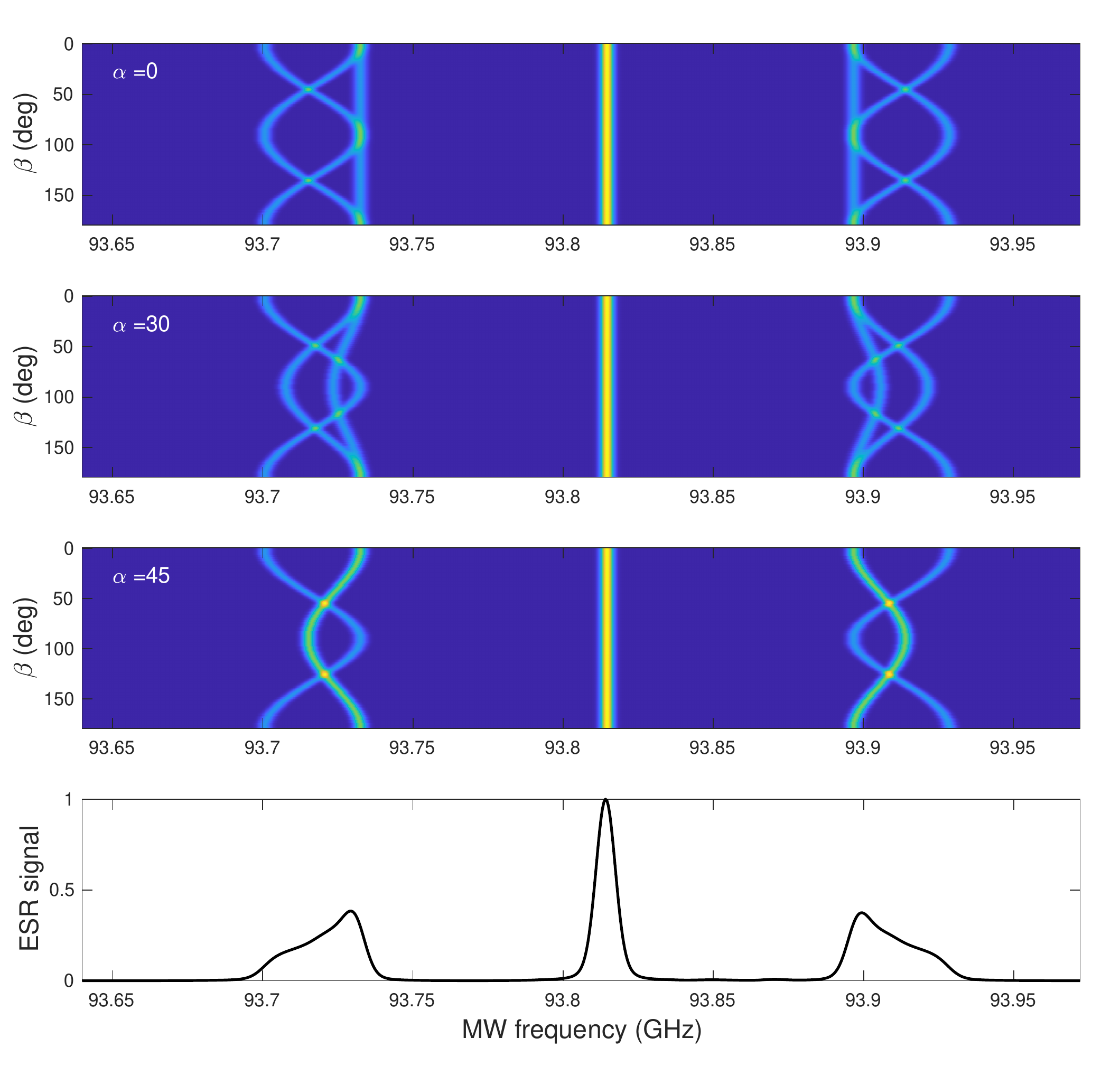}
\caption{Easy spin simulations of the EPR spectrum of P1 centers in diamond as a function of crystal orientation.  Depending on the orientation, there are up to 4 distinct orientations of the P1 center with respect to the magnetic field for a single crystal orientation.
The Euler angles $(\alpha, \beta, \gamma)$ indicate the orientation of the lattice with the external magnetic field.  We have set $\gamma=\pi/4$ and the top three subplots indicate rotation patterns as a function of $\beta$ with $\alpha = 0, \pi/6$ and $\pi/4$.  The lowest curve shows the simulated powder spectrum \label{fig:EPR}}
\end{figure}

Experimental results from a second crystal orientation exhibiting the same DNP mechanisms are shown in Section~\ref{SISec:singleXtal} of the SI.  While no direct CE is observed at these crystal orientations, this will not necessarily be true for all orientations.  In order for a diamond crystal to satisfy the conditions for CE enhancement at a given orientation, there should be two EPR lines separated by the nuclear Larmor frequency.  In order to understand the DNP spectrum from the powder, we studied how the EPR spectrum from the diamond sample changes with orientation in the external field.  Figure \ref{fig:EPR} shows EasySpin simulations of how the single crystal spectra change as the crystal is rotated about different axes, as well as the powder averaged spectrum \cite{stoll2006}. There are a number of orientations at which the separation between two of the lines in the $m_I = \pm 1$ manifold is on the order of the $^{13}$C Larmor frequency. Thus two adjacent P1 centers - with different relative orientations - can undergo a dipolar-driven mutual spin-flip that would lead to $^{13}$C DNP. (Other CE conditions may be fulfilled at other fields, such as those described by Bretschneider et al.\cite{Bretschneider2016}). Note that the addition of electron-electron dipolar interactions will also cause a broadening of the EPR lines, making the CE condition easier to fulfill. 

\begin{figure}[t]
\includegraphics[width=0.48\textwidth]{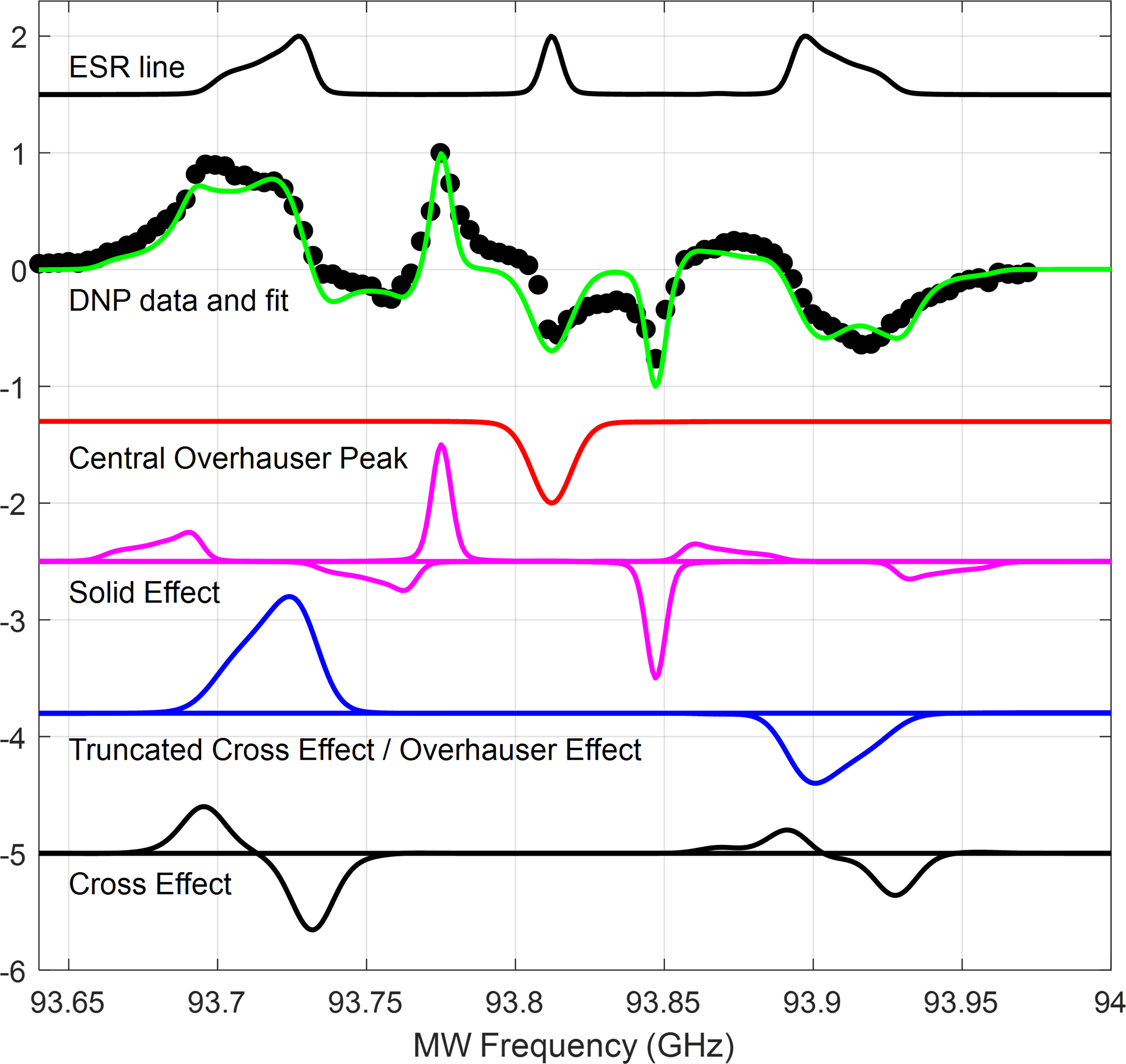}
\caption{Fit of experimental constant-frequency DNP spectrum (black circles) of the diamond powder sample using a sum (green line) of the OE (red line), SE (magenta line), the truncated CE (blue line), and the CE (black line). The EPR line plotted above includes line broadening of 10 MHz of the outer EPR lines. \label{fig:DNP_spectrum_fit}}
\end{figure}

Section~\ref{SIsec:eNsystem} of the SI provides a detailed description of an e-N-C spin system, and full quantum mechanical simulations of the DNP spectra for both $^{13}$C and $^{14}$N, including relaxation. These simulations reveal that, in general, the $^{13}$C-SE-DNP enhancement occurs within a $^{14}$N manifold (i.e.\ the nitrogen spin state does not affect the carbon DNP). For the simulations we choose to concentrate on $^{13}$C-SE-DNP due to its relative simplicity to understand.


\subsection{Powder}
\noindent Figure \ref{fig:DNP_spectrum_fit} shows a fit of the DNP spectrum of the powder sample. The components of the SE, OE, tCE and CE with their respective intensities are also plotted. As in the single crystal case, the center of the DNP spectrum can be fit using a combination of the SE and the OE giving negative enhancement. Here the outer lines are best fit using a combination of the CE and the OE or truncated CE.

Figure~\ref{fig:DNP_spectrum}(b) showed that chirped DNP  significantly improved the enhancements for the two outer nuclear spin manifolds ($m_I = \pm 1$), and blurred some of the sharper features associated with the central manifold ($m_I = 0$) \cite{hovav2014dynamic,Guy2017,Shimon-2019}.  The enhancement observed under modulation could either be due to the cross effect \cite{hovav2014dynamic,Guy2017} or due to the introduction of a new mechanism -- the integrated solid effect \cite{Henstra-1988}. Given the low microwave power and the rapid 5 kHz modulation rate, significantly faster than the electron T$_1$, 
it is unlikely that the experiment satisfies the conditions necessary for achieving the integrated solid effect via frequency modulation \cite{Can2017,Can2018}.  Additionally the effect is not observed in the central ($m_I = 0$) manifold.  Thus the additional enhancement from modulation is believed to be due to the CE.

We have also characterized the build-up times and the power-dependence of the DNP spectrum (see Sections~\ref{SISec:DNP-build-up} and \ref{SISec:power} of the SI).  Both sets of data suggest that the observed DNP enhancements are limited by the available microwave power.

The plurality of DNP mechanisms observed is more complex than in previous high-field DNP experiments with P1 centers.  In a diamond micropowder sample with a P1 concentration less than 100 ppm,
Bretschneider {\em et al}.\ observed three SE-DNP lines, one from each EPR line, and a single positive OE line from the central EPR line under static DNP conditions at W-band (94 GHz) and low temperature (1.5-100 K) \cite{Bretschneider2016}.
Similarly, Kwiatkowski {\em et al}.\ observed static DNP from P1 centers at 3.5 K at both 94 GHz and 196 GHz in nanodiamonds via a combination of the SE and a small positive OE  \cite{Kwiatkowski2018}. The positive OE enhancements measured at low temperature are {\em opposite} to the negative OE we observe at room temperature.  The origin of this difference is still unknown.

\section{Sample Heterogeneity}  \label{sec:heterogeneity}

To our knowledge this is the first case in which the OE, SE, CE and truncated CE have been observed in a single material system under the same experimental conditions.   This is likely due to the heterogeneity in the distribution of nitrogen in these Type Ib HPHT diamond samples.  

Li et al.\ recently used double electron-electron resonance (DEER) experiments to show that single crystal HPHT diamond samples -- similar to those studied here -- show significant spatial variations in their P1 concentrations \cite{Li-2021}.  One of the samples they studied was measured to have local P1 concentrations ranging from 13 ppm to 322 ppm in different regions.  

\begin{figure}[t]
\includegraphics[width=0.48\textwidth]{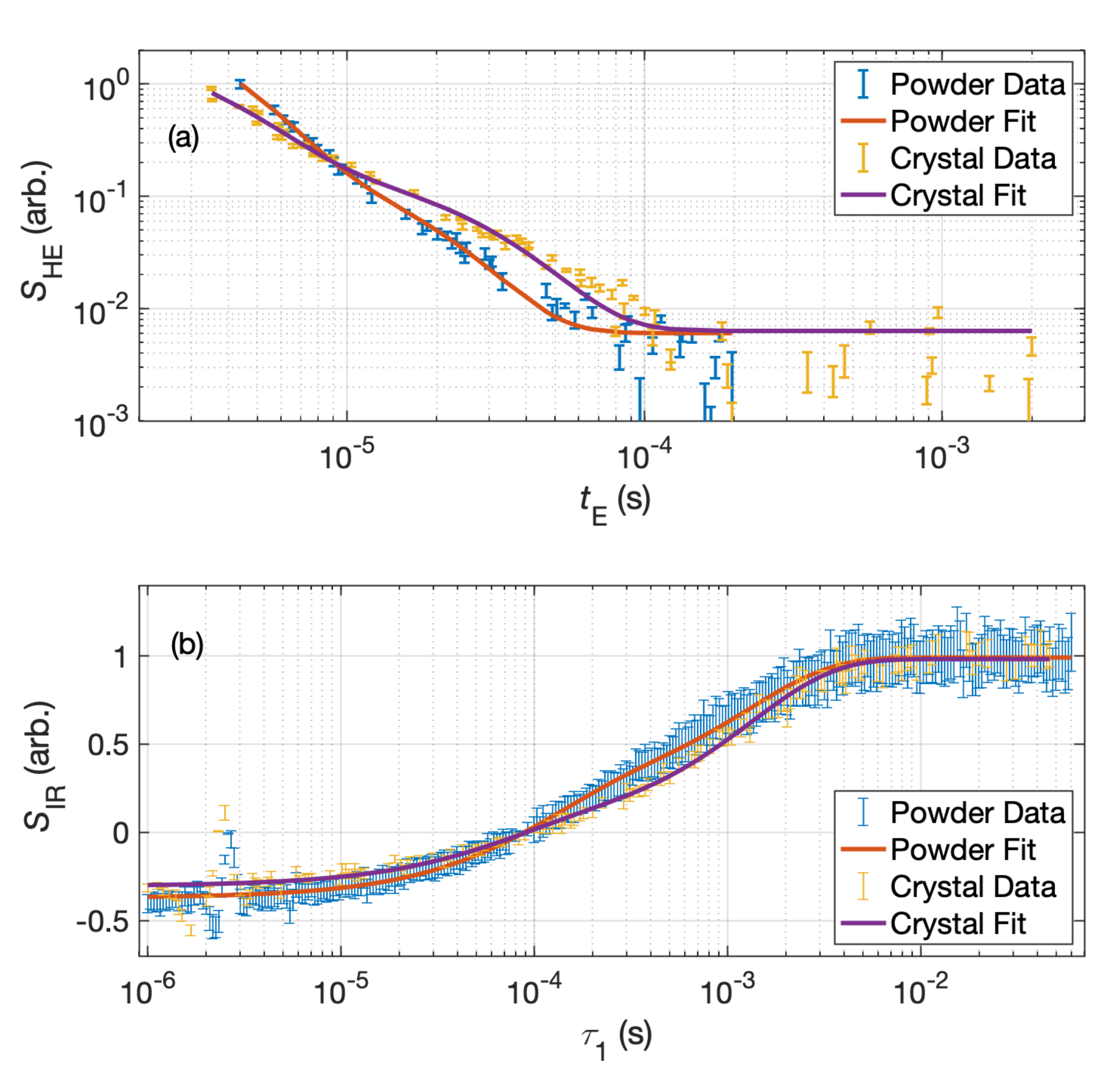}
\caption{Results from low-field pulse-EPR experiments. The fit parameters are listed in Tables \ref{tbl:HahnEchoT2e} \& \ref{tbl:InversionRecT1e} in the Materials and Methods. Experimental details can be found in Section \ref{subsec:MatMethPepr} (Materials and Methods) and Section \ref{SISec:PeprDetails} of the SI. (a) Hahn-echo decays showing bi-exponential behavior, which suggests there are at least two types of local environment with different P1 concentrations. (b) Inversion recovery experiment indicating fast and slow electronic T$_1$ relaxations. The field dependence is expected to be weak at room temperature as it is likely to be dominated by two-phonon Raman transitions. \label{fig:EPR_T1T2_fit} }
\end{figure}

We measured the T$_1$ and T$_2$ of the P1 center in both our powder sample and a similar single crystal sample to that used for DNP.  The  2.5 GHz pulsed ESR setup is described in Section \ref{SISec:PeprDetails} of the SI.  Figure \ref{fig:EPR_T1T2_fit} shows that for both the Hahn-echo and the inversion recovery experiments, bi-exponential relaxation best fits the data. The Hahn-echo data are fit with time constants of about 2 $\mu$s and 10 $\mu$s corresponding to $\sim 80$ ppm and $\sim 16$ ppm using the relation $1/T_2$ ($\mu$s$^{-1}$) = $\frac{1}{160}$ ($\mu$s$^{-1}$ppm$^{-1}$) $\times$ [P$_1$] (ppm) \cite{Bauch-2020}. Note that it is difficult to measure T$_2$ values below 1 $\mu$s on our system due to instrument limitations, so there could potentially be pools with even higher local P$_1$ concentrations present.  The two T$_1$ time constants were about 100 $\mu$s and 1.4 ms.

This heterogeneity is key to understanding how the different mechanisms observed above appear to co-exist in the same sample. In regions with low P1 concentration, the relatively isolated defects exhibit DNP via the solid effect.  As the P1 concentration increases, we see the appearance of the CE in those crystallites where the orientation permits the cross effect condition to be satisfied. Some of the local P1 clusters become fast-relaxing sites that are then responsible for the appearance of the truncated cross effect.  As these cluster resonances can be fairly broad \cite{Slichter-1955}, it is possible to satisfy the truncated cross effect conditions at most of the crystal orientations.  If the P$_1$ in the clusters are close enough for exchange interactions to become important, this could also explain the origin of the Overhauser effect.  Overhauser DNP has previously been observed in graphite \cite{Abragam-1958} as well as in exchange-coupled donor clusters in silicon \cite{Dementyev-2011}. However, a more systematic study of the DNP spectrum as a function of magnetic field and temperature is needed to uniquely identify the underlying mechanisms for the observed Overhauser effects, especially in light of the different signs observed for this effect in different experimental regimes.

\section{Conclusions}
\label{sec:conclusions}

In summary, we have shown substantial $^{13}$C-DNP enhancement ($>100$) with P1 centers in diamonds at room temperature.  We have also shown that the DNP enhancement proceeds via a complex combination of SE, OE, CE and tCE DNP mechanisms, all within the same sample. The DNP enhancements observed are limited by the available power and could be improved by going to higher power and potentially higher magnetic field.  Additionally, it might be possible to hyperpolarize the P1 centers via their interactions with adjacent optically-polarized NV centers which could dramatically improve DNP enhancements even further \cite{Loretz-2017, PhysRevB.96.220407}.  

Using DNP to increase the sensitivity of NMR has the potential to significantly expand the application of this versatile spectroscopic technique to ever-smaller samples.  This technique will also be useful for enhancement of NMR signals in NV-detected NMR at a high field \cite{Fortman-2021} and be potentially applicable to NV-detected NMR of external spins.  
Hyperpolarized NMR of Nano- and micro-diamond tracers could prove valuable in both biomedical applications as well as in fluid engineering.  It should also be possible to transfer the hyperpolarization to spins external to the diamond via spin diffusion \cite{vanderWel-2006} or cross-polarization \cite{Shagieva2018}. DNP with diamond chips would enable the use of magnetic resonance to study oriented low-dimensional systems such as thin films and 2D materials like graphene and its functional derivatives, transition-metal dichalcogenides, and 2D conductive metal-organic-frameworks.  These materials are increasingly being used as catalysts and chemiresistive devices, and may hold promise for the development of novel quantum materials.

\section{Materials and Methods}

\subsection{High Field CW-EPR Experiment}

\noindent Room-temperature continuous-wave (cw) electron paramagnetic resonance (EPR) lines were measured at 230 GHz. Field modulation frequency is 20 kHz and the modulation strength is ~0.02mT.

\subsection{Low Field Pulse-EPR Experiments}
\label{subsec:MatMethPepr}

\noindent Our lab-built pulse-EPR spectrometer is designed to operate at 2.5 GHz and can output up to 1 W of power. In Figure \ref{fig:EPR_T1T2_fit}, data are shown for (a) Hahn-echo and (b) inversion recovery experiments conducted on the powder sample and on a single crystal macle-cut Element 6 HPHT diamond with $\textbf{B}_0$ parallel to $[111]$. The data were collected with the external magnetic field tuned to the $m_I = 0$ manifold of the EPR line. For the Hahn-echo experiment, the sequence is $\pi/2 - \tau - \pi - \tau - \textrm{echo}$, and the interpulse delay $\tau$ is swept. The magnitudes of the echoes are plotted as a function of the time at which the echo comes into focus. For both the powder and the single crystal diamond, the Hahn echo magnitudes, $S_\text{HE}(t_\text{E})$ are fit to a bi-exponential decay equation,
\begin{equation}
    S_\text{HE}(t_\text{E}) = M_1e^{-t_\text{E}/T_2^1} + M_2e^{-t_\text{E}/T_2^2} + A,
\end{equation}
where $t_\text{E} = 2\tau + t_{\pi/2}/2 + t_{\pi}$ is the cumulative time that spins are undergoing $T_2$ relaxation, $M_1$ and $T_2^1$ are the amplitude and transverse relaxation time characteristic of electrons in pool 1, $M_2$ and $T_2^2$ are the likewise variables for pool 2, and $A$ is an offset that is necessarily non-zero since the echo magnitudes are used in fitting. The fit parameter values are given in Table \ref{tbl:HahnEchoT2e}.
\begin{table}[hb]
\begin{tabular}{c|c|c|c|c|c|}
\cline{2-6}
                                                                               & $M_1$    & $T_2^1$ ($\mu$s) & $M_2$     & $T_2^2$ ($\mu$s) & A          \\[1.5pt] \hline
\multicolumn{1}{|c|}{Powder}                                                   & $8(1)$   & $1.9(1)$         & $0.30(9)$ & $10(2)$          & $0.006(4)$ \\ \hline
\multicolumn{1}{|c|}{\begin{tabular}[c]{@{}c@{}}Single\\ Crystal\end{tabular}} & $3.2(9)$ & $2.2(5)$         & $0.2(1)$  & $18(8)$          & $0.01(1)$  \\ \hline
\end{tabular}
\caption{Fit parameters for Hahn-echo low-field pulse-EPR experiments. Uncertainties indicate 95\% confidence intervals.
\label{tbl:HahnEchoT2e}}
\end{table}

For the inversion recovery experiments, the sequence $\pi - \tau_1 - \pi/2 \tau_2 - \pi - \tau_3 - \textrm{echo}$ inverts the electron spin polarization with the first $\pi$ pulse, then after a recovery time $\tau_1$, reads out the polarization with a standard Hahn-echo (note, $\tau_2 = \tau_3$). By sweeping $\tau_1$ from short to long values, the echo signal transitions from a maximally inverted signal to one that is fully recovered via $T_1$ relaxation processes. The echo amplitudes $S_\text{IR}(\tau_1)$ are well fit with a bi-exponential asymptotic recovery equation,
\begin{multline}
    S_\text{IR}(\tau_1) = M_1\left(1-2\sin^2(\alpha/2)e^{-\tau_1/T_1^1}\right) +\\ M_2\left(1-2\sin^2(\alpha/2)e^{-\tau_1/T_1^2}\right),
\end{multline}
where $M_1$ and $T_1^1$ are the amplitude and longitudinal relaxation time characteristic of electrons in pool 1, $M_2$ and $T_1^2$ are the likewise variables for pool 2, and $\alpha$ is an angle whose deviation from $\pi$ indicates imperfect inversion of the electron spin population. The values of the fit parameters are given in Table \ref{tbl:InversionRecT1e}.

\begin{table}[ht]
\begin{tabular}{c|c|c|c|c|c|}
\cline{2-6}
                                                                               & $M_1$     & $T_1^1$ (ms) & $M_2$     & $T_1^2$ (ms) & $\alpha$ (rad) \\[1.5pt] \hline
\multicolumn{1}{|c|}{Powder}                                                   & $0.41(4)$ & $0.11(2)$    & $0.58(4)$ & $1.3(1)$     & $4.33(1)$      \\ \hline
\multicolumn{1}{|c|}{\begin{tabular}[c]{@{}c@{}}Single\\ Crystal\end{tabular}} & $0.25(5)$ & $0.07(3)$    & $0.74(5)$ & $1.3(2)$     & $4.40(3)$      \\ \hline
\end{tabular}
\caption{Fit parameters for inversion recovery low-field pulse-EPR experiments. Uncertainties indicate 95\% confidence intervals.
\label{tbl:InversionRecT1e}}
\end{table}

\subsection{Referencing the $^{13}$C-NMR Spectrum}
\noindent The $^{13}$C-NMR spectrum was referenced to adamantane, at 37$\pm$1 ppm using glycerol-d5 as a secondary reference, with two peaks at 64$\pm$1  and 72$\pm$1 ppm. 

\subsection{DNP Spectrometer}
\noindent The DNP experiments were performed on a homebuilt DNP spectrometer, at a field of 3.34 T, corresponding to an electron Larmor frequency of 94 GHz, a $^{1}$H Larmor frequency of 142 MHz and a $^{13}$C Larmor frequency of 35.8 MHz. \cite{Guy-2015} All experiments were performed at room temperature. The NMR pulses and detection were controlled with a Bruker Avance AQX spectrometer.

\subsubsection{MW Irradiation} 

\begin{figure}
\includegraphics[width=0.5\textwidth]{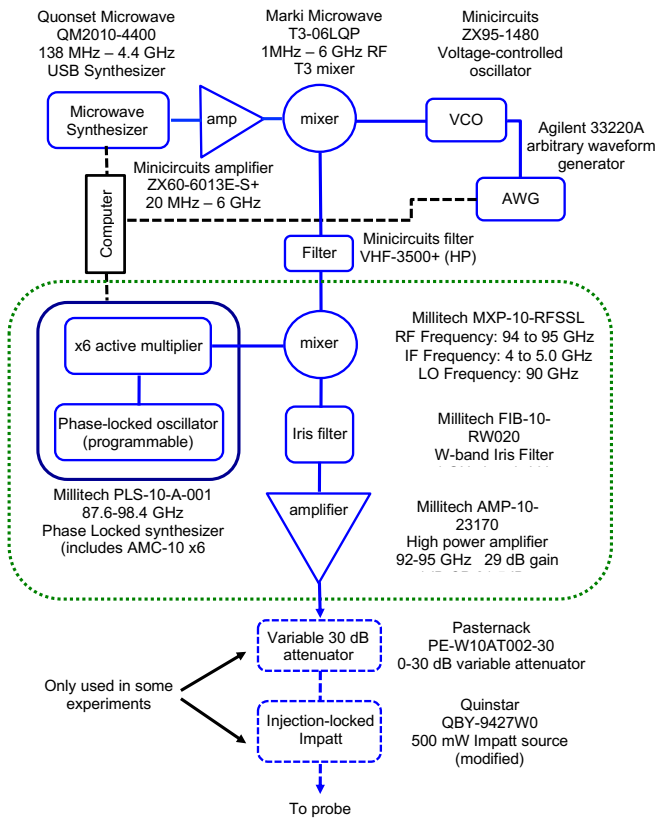}
\caption{Schematic of the millimater wave source used in these experiments - adapted from \cite{Guy-2015}.}
\label{fig:MW}
\end{figure}

\noindent An Agilent 33220A arbitrary waveform generator (AWG) was used to create the waveforms used for continuous wave (CW) MW irradiation. The AWG output was set to 2.5 V, which was input into a Mini Circuits voltage controlled oscillator ZX95-1480+ (VCO) for CW or chirped excitation around a central frequency of 1.012 GHz. The 1.012 GHz output of the VCO was first mixed (Marki T3 mixer) with a variable frequency signal from the ($\sim 4$ GHz) Quonset Microwave QM2010-4400 source to yield sidebands at $\sim 5$ GHz and $\sim 3$ GHz. The 3 GHz component mixer was filtered out using a Mini Circuits high-pass filter VHF-3500+ that only allows frequencies above 3.9 GHz to pass through. The variable frequency signal around 5 GHz was then mixed with an 88.56 GHz phase-locked oscillator source to reach 94 GHz, and filtered again to remove the low frequency sideband, using an iris filter as shown in Figure~\ref{fig:MW}. This is a modification of the Millitech MW bridge previously described in \cite{Guy-2015}.  In order to increase the available millimeter wave power we removed the voltage controlled attenuator (1.5 dB insertion loss) and the directional coupler (1.8 dB insertion loss) to get a maximum output power of 23.8 dBm (240 mW).

In order to vary the power below 240 mW we connected a variable 30 dB attenuator to the output to the source.  In order to run the experiment at higher power, we modified a commercial 500 mW injection-locked 94 GHz source from Quinstar.  We removed the varactor-tuned Gunn diode orignally used for the injection locking from the source and instead used the attenuated output of the Millitech source.  We were unable to verify if the efficiency of injection-locking was uniform across the entire sweep bandwidth.

\subsection{$^{13}$C-NMR and DNP Experiments}

\noindent The following settings were used for all the experiments, unless otherwise noted: The DNP enhanced NMR signal was recorded with a 90-acquire pulse sequence, using a 10 $\mu$s $\pi$/2 pulse. Eight step phase cycling was used in all cases. The experiments began with a train of 100 30 ms saturation pulses separated by 20 $\mu$s.

The $^{13}$C relaxation time T$_{1}$ (Section~\ref{SISec:thermal-build-up} of the SI) was recorded using saturation-recovery experiments, stepping the recovery delay and then recording the signal, with no MW irradiation applied. The DNP enhancement buildup time T$_{bu}$ was recorded in the same manner as T$_{1}$ but with MW irradiation.

DNP spectra were recorded by keeping the 1.012 GHz VCO settings constant, and stepping the 4 GHz source in order to cover a range of 93.63 GHz to 93.972 GHz. Data were typically taken over 102 evenly spaced frequencies points.

\subsection{NMR Data Processing} 

\noindent Data processing was performed in MATLAB using custom scripts. 3 points of left-shift were used in all cases to remove switching signal from opening of the receiver. The data was then baseline corrected, phase corrected and 300 Hz exponential line broadening was applied. For DNP enhancement calculations we divide the integrated intensity of the MW-on signal from the MW-off signal (such that no enhancement is equal to 1). The MW-off signal was measured using 128 scans, except at the longest build-up times when 64 scans were used.

The reported signal intensities correspond to an average over 21 points around the peak of the phased absorptive signal in the frequency domain. The reported uncertainties in the NMR signal were calculated using the standard deviation (201 point interval) from a signal-free region of the NMR spectrum.  

The T$_{1}$ and T$_{bu}$ curves were fit using a biexponential function, using the method of a non-linear least squares. The fitting function returns a 95 \% confidence interval which was converted to a variation of $\pm 2\sigma$ assuming a normal distribution.

The authors thank Hailey Mullen for helping acquire the DNP spectrum of the diamond crystal and Element 6 for the donation of the powder samples used in these experiments.  We thank Professor David Cory (Waterloo) for the 500 mW Impatt source.  This work was partially supported by funding from the National Science Foundation under cooperative agreement under grants OIA-1921199.  ST acknowledges support from the US National Science Foundation (DMR-1508661, CHE-1611134, and CHE-2004252 with partial co-funding from the Quantum Information Science program in the Division of Physics).

The data can be provided by the authors pending scientific review and a completed material transfer agreement. Requests for the data should be submitted to: Daphna Shimon or Chandrasekhar Ramanathan.

\bibliography{diamond_refs.bib}

\bigskip

\begin{widetext}
\newpage

\section*{Supplementary Information: Large Room Temperature Bulk DNP of $^{13}$C via P1 Centers in Diamond}
\setcounter{page}{1}

Daphna Shimon$^{a1}$, Kelly A. Cantwell$^{b}$, Linta Joseph$^{b}$, Ethan Q. Williams$^{b}$, Zaili Peng$^{c}$, Susumu Takahashi$^{c}$, Chandrasekhar Ramanathan$^{b2}$

\noindent a: Institute of Chemistry, The Hebrew University of Jerusalem, Edmond J. Safra, Givat Ram, Jerusalem, Israel.

\noindent b: Department of Physics and Astronomy, Dartmouth College, Hanover, NH 03755, U.S.A.

\noindent c: Department of Chemistry, University of Southern California, Los Angeles, California 90089, U.S.A.

\noindent 1: daphna.shimon@mail.huji.ac.il

\noindent 2: chandrasekhar.ramanathan@dartmouth.edu
\bigskip

\end{widetext}

\setcounter{section}{0}

\section{$^{13}$C decoherence time}
\label{SISec:decoherence}

The NMR peak has a width of 1.12 kHz. This resonance shows significant inhomogeneous line-broadening and it is possible to detect several hundred echoes in a pulsed spin-lock experiment (using a train of $\pi/2$ pulses \cite{beatrez2021}) as shown in Figure~\ref{fig:NMR_spectrum}(b). Stroboscopic detection of multiple echoes would allow a significant {\em additional} improvement in the signal-to-noise ration (SNR) obtained.  The decay times measured are significantly shorter than those recently measured by Beatrez {\em et al}.\ \cite{beatrez2021}, potentially due to the microwaves being continually on during the experiment.

\begin{figure}
\includegraphics[width=0.49\textwidth]{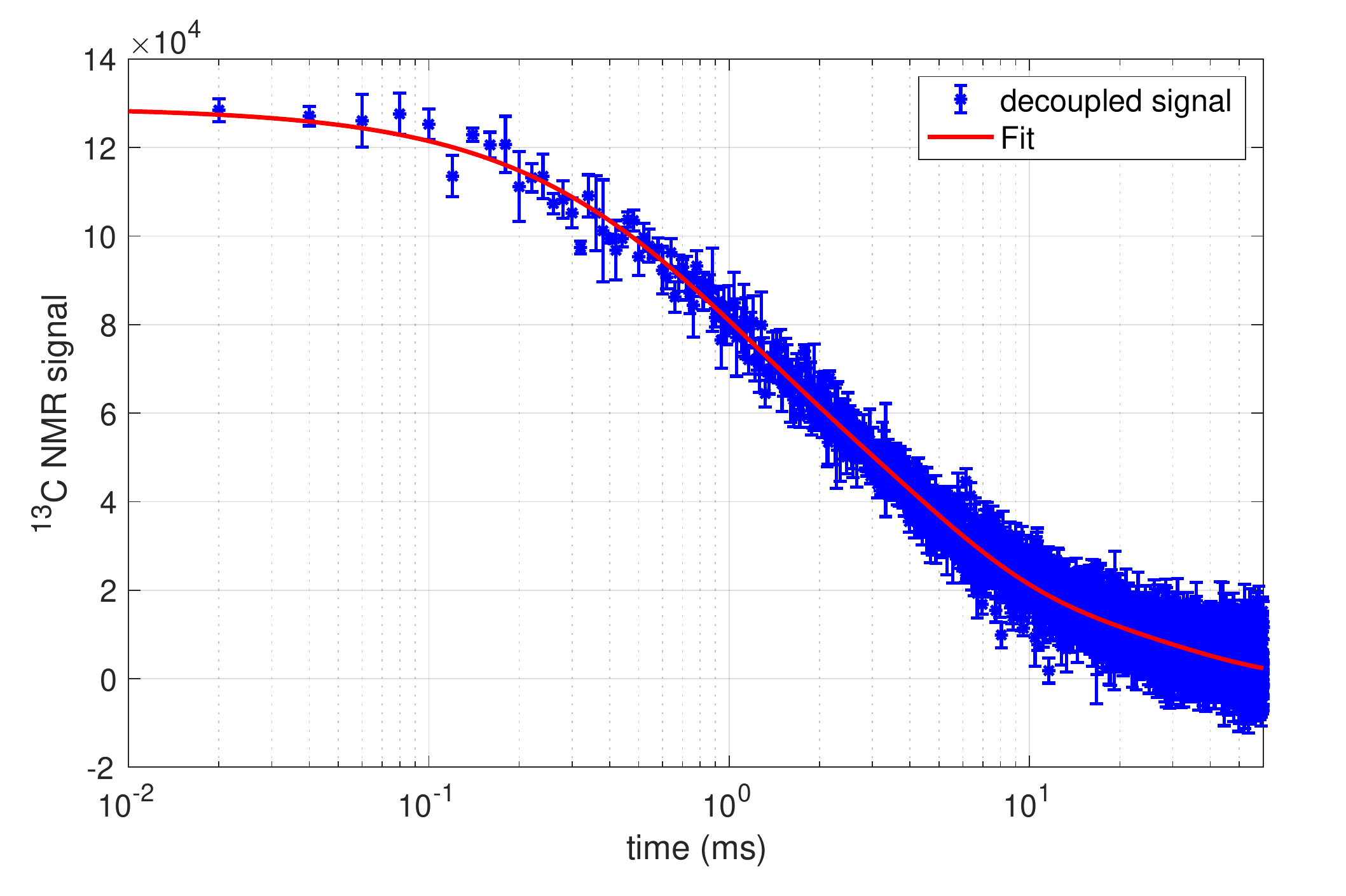}
\caption{The decay of a pulsed spin-locking echo train with a 10 $\mu$s $\tau$ spacing  following hyperpolarization for 60 s. The data were collected in a two-dimensional experiment. The best fit is obtained with a 3-component fit with time constants of $25.5 \pm 1.5$ ms, $3.8 \pm 0.4$ ms and $0.69 \pm 0.13$ ms with relative intensities of 0.19, 0.47 and 0.33.}
 \label{fig:decoherance}
\end{figure}

\section{DNP Enhancement of the Single Crystal}
\label{SISec:singleXtal}
\begin{figure}
\includegraphics[width=0.49\textwidth]{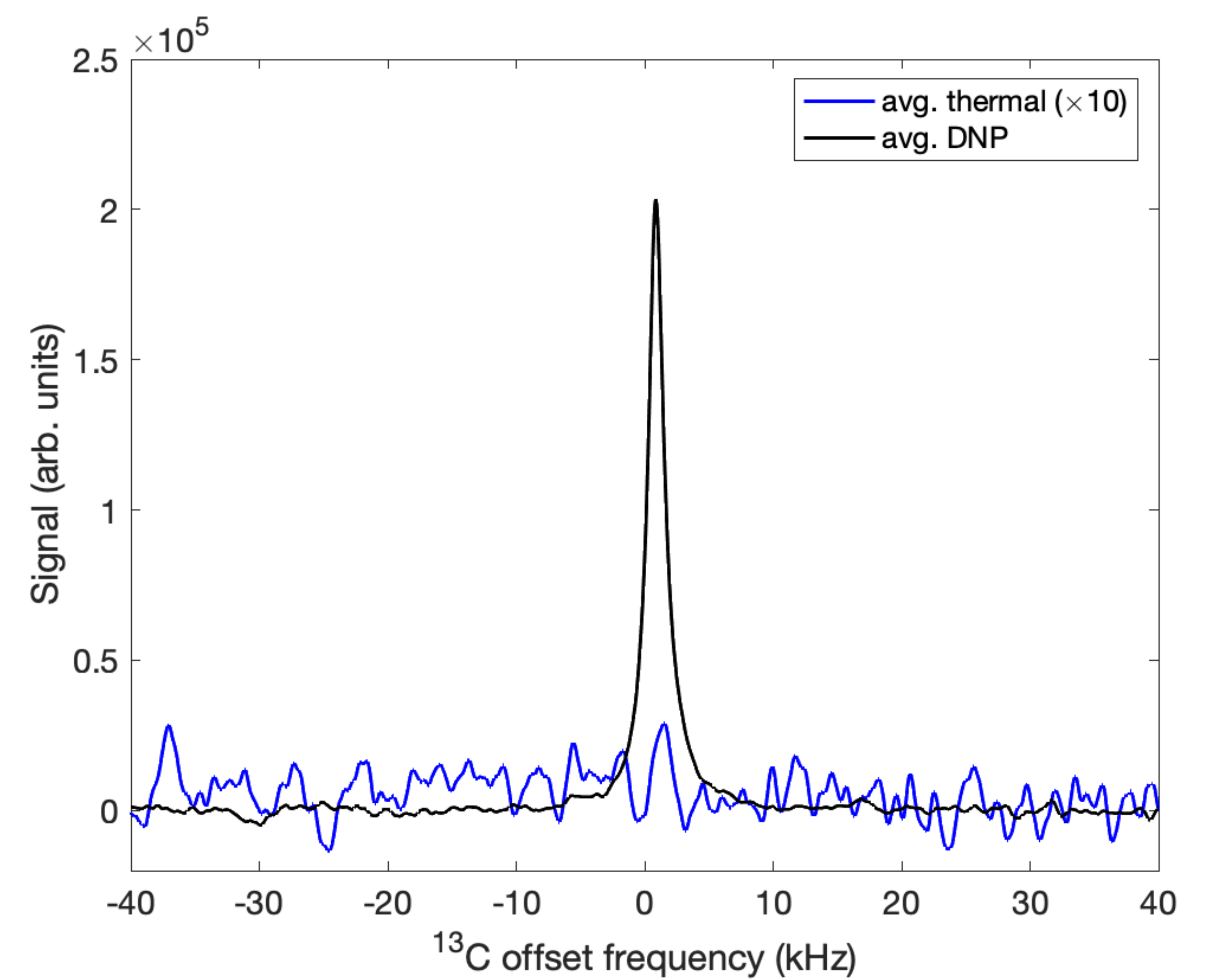}
\caption{Thermal and DNP-enhanced $^{13}$C NMR spectra from the single-crystal diamond sample.  The thermal signal is an average of 800 scans and has been vertically scaled by a factor of 10 to better visualize the signal.  The DNP signal is an average of 8 scans and is acquired with the crystal at the orientation shown in Figure~\ref{fig:DNP_spectrum_xtal_fit}.
 \label{fig:xtal-enhancement}}
\end{figure}

\noindent We tried to measure the thermal equilibrium $^{13}$C signal from the single crystal diamond sample.  Figure~\ref{fig:xtal-enhancement} shows average of 800 scans of the thermal signal measured with a relaxation delay of 500 s as well as the average of 8 scans of the DNP signal measured at 93.77 GHz with the same build-up of 500 s with about 240 mW of microwave power.  Since we are unable to see the thermal $^{13}$C signal, we use the SNR of the DNP signal to estimate a minimum DNP enhancement of 180 for the single crystal sample.  Note the thermal signal was acquired with a receiver gain that was 4 times larger than that used in the DNP experiment.

\begin{figure}[htb]
\includegraphics[width=0.48\textwidth]{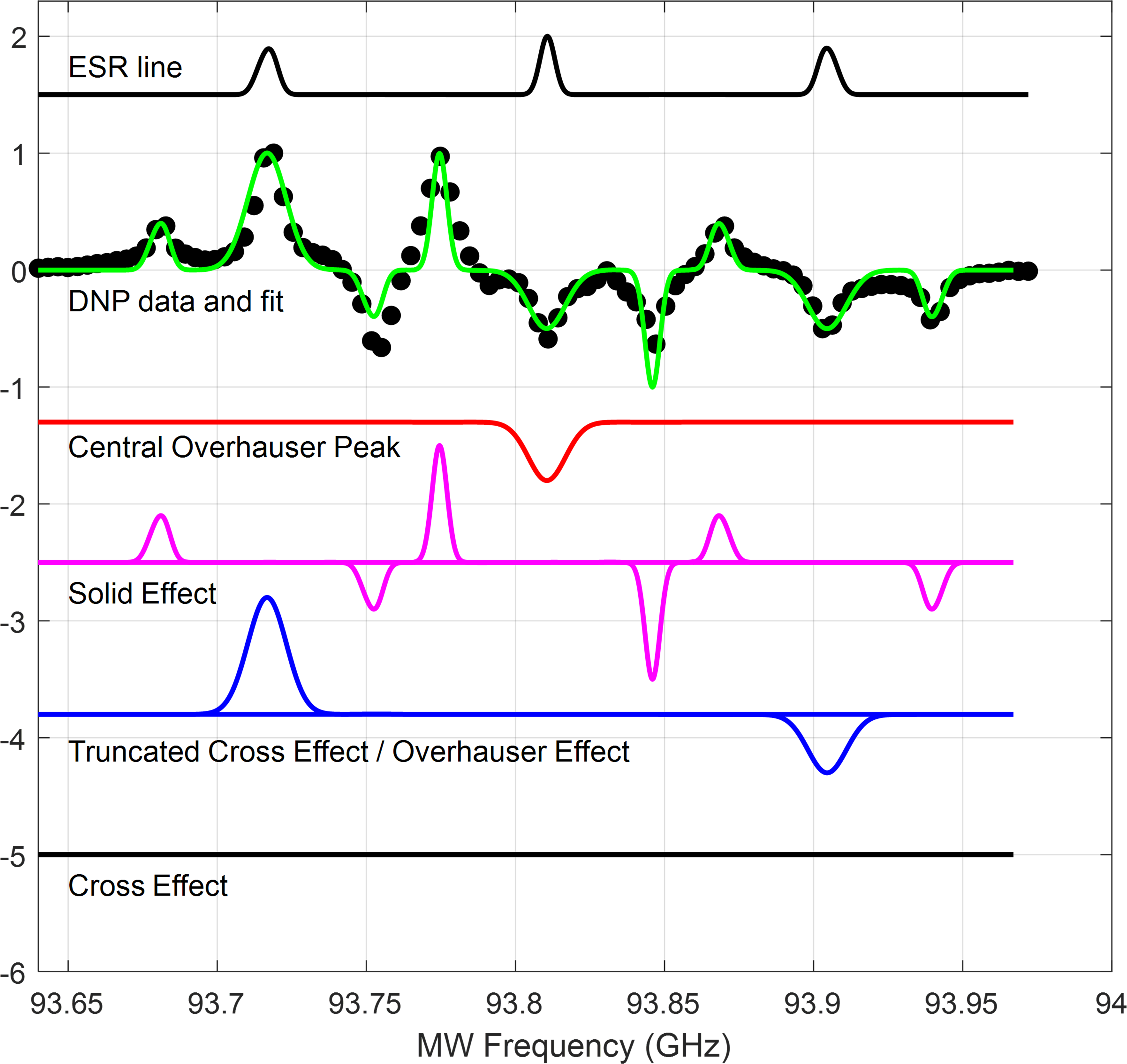}
\caption{Fit of experimental constant-frequency DNP spectrum (black circles) for a single-crystal diamond sample.  The fit uses a sum (green line) of the OE (red line), SE (magenta line), the truncated CE (blue line), and the CE (black line). The EPR line plotted above includes line broadening of 10 MHz of the outer EPR lines. The single crystal EPR line was simulated with EasySpin using the Euler angles
$\alpha=136.8000^{\circ}$, $\beta=129.7918^{\circ}$, $\gamma=309.6000^{\circ}$.
\label{fig:DNP_spectrum_xtal_fit_2}}
\end{figure}

In Figure \ref{fig:DNP_spectrum_xtal_fit_2} we show a fit of a single crystal at a second orientation, using the method described in the main text, and below.

\section{Simulating the EPR Spectrum}
\label{SISec:simulateEPR}
\noindent Given the concentration of P1 centers it is also important to account for magnetic dipolar interactions between electron spins in the crystal. Without knowing the specific distribution of P1 centers in the system we assume here that the P1 centers are uniformly distributed.  We used a Monte Carlo simulation to randomly seed a diamond lattice with a 100 ppm concentration of P1 centers and estimated the pairwise distribution of dipolar couplings in the sample by averaging over a large number of configurations.  

\begin{figure}
\includegraphics[width=0.5\textwidth]{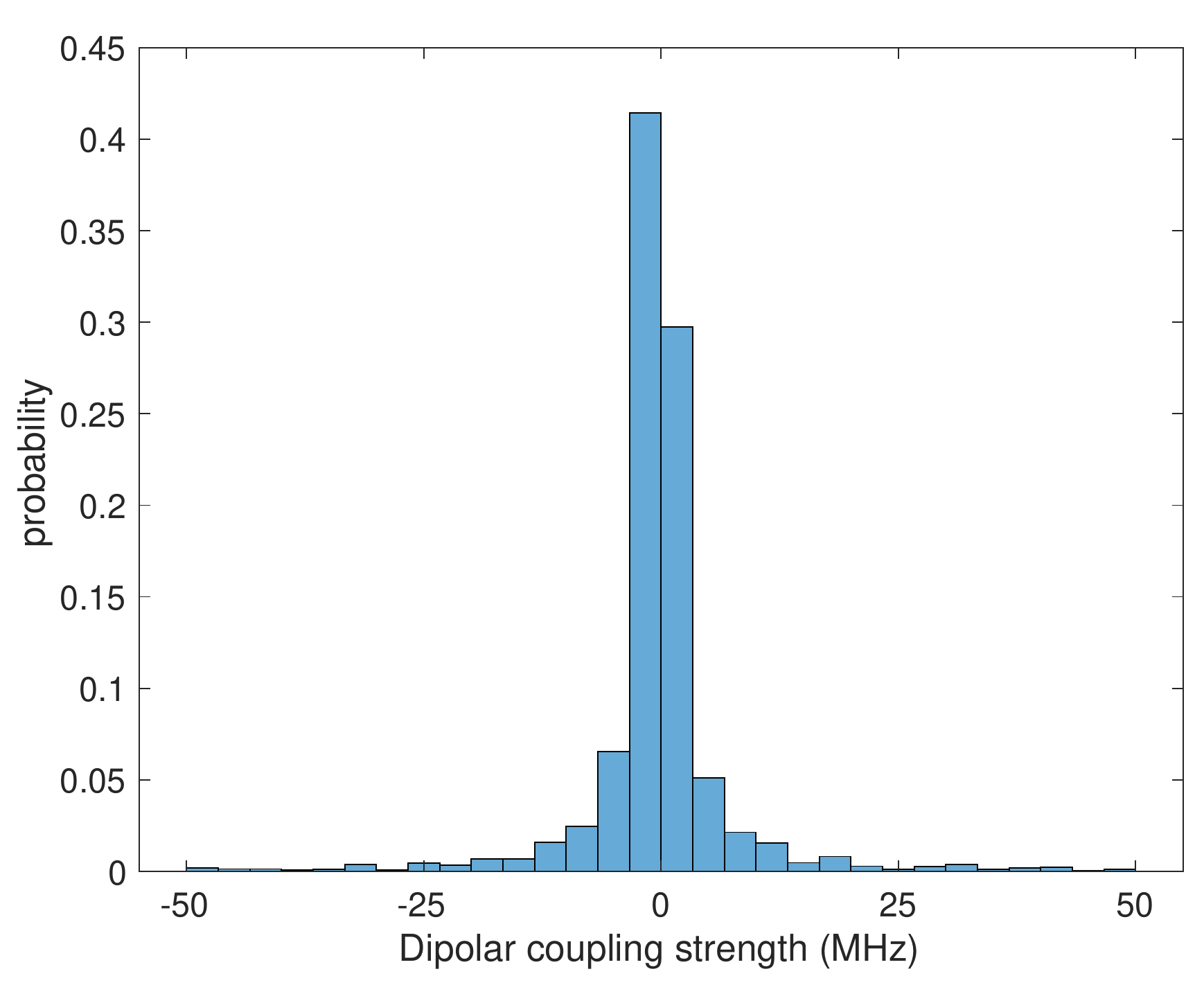}
\caption{Simulated distribution of pairwise dipolar couplings for a 100 ppm P1 concentration diamond sample.}
\label{fig:couplings}
\end{figure}

We used a lattice with $15 \times 15 \times 15$ unit cells, each with 2 atoms per unit cell and randomly seeded the sites with P1 centers using a binomial probability determined by the concentration.  The distribution of pairwise P1 dipolar interactions was then calculated and stored.  We averaged this distribution over 10,000 random lattice configurations to estimate the distribution of pairwise dipolar couplings in the system as shown in Figure~\ref{fig:couplings}.

We then used EasySpin package \cite{stoll2006} to perform quantum mechanical simulations of pairs of P1 centers .  As noted earlier, each P1 center was modeled as an e-$^{14}$N system, with an isotropic g=2.0024, hyperfine coupling strengths with the $^{14}$N nucleus of $A_{x}^N=A_{y}^N$ = 82 MHz, A$_{z}^N$ = 114 MHz \cite{Loubser-1978}. The $^{14}$N nuclear quadrupolar interaction was neglected. EasySpin automatically computes the powder average of the spectrum and uses the 227 (F d -3 m) space group symmetry of the diamond lattice.

For the single crystal simulation shown in Figure \ref{fig:DNP_spectrum_xtal_fit}, we simulated a single crystal orientation with Euler angles $\alpha=54.0013^{\circ}$, $\beta=134.4274^{\circ}$, $\gamma=18.0023^{\circ}$, and the same space group, isotropic g-value and hyperfine interactions described in the previous paragraph. For the single crystal simulation shown in Figure \ref{fig:DNP_spectrum_xtal_fit_2} we simulated a single crystal orientation with Euler angles
$\alpha=136.8000^{\circ}$, $\beta=129.7918^{\circ}$, $\gamma=309.6000^{\circ}$.

\section{Fitting the Experimental DNP Spectrum}
\label{SISec:FittingDNP}
\noindent In the main text, we fit the DNP spectrum using a crude method of convoluting the EPR line with delta functions to form the basic shapes for the solid effect (SE), Overhauser effect (OE) (same shape  used for tCE) and cross effect (CE) DNP mechanisms, and changing their amplitude to achieve the best agreement with the experimental spectrum. This method was previously described in Banerjee et al. \cite{Banerjee2013}.

In order to do this, the EPR line, $g(\omega_{\text{MW}})$,  was split into three ranges:
\begin{itemize}
\item 
The low frequency line: $g_1(\omega_{\text{MW}})$ for $93.56\leq\omega_{\text{MW}}\leq93.7740$ GHz
\item 
The central line: $g_2(\omega_{\text{MW}})$ for $93.7740<\omega_{\text{MW}}<93.8590$ GHz
\item
The high frequency line: $g_3(\omega_{\text{MW}})$ for $93.8590\leq\omega_{\text{MW}}\leq94.0610$ GHz
\end{itemize}

The EPR line used for the powder DNP spectrum was simulated using using a e-$^{14}$N e-$^{14}$N spin system, averaged over 100 of the most probable e-e dipolar interactions between the two systems (see Figure~\ref{fig:couplings}). This averaging caused a slight broadening of the two external EPR lines (as explained above). For the single crystal DNP spectrum, we simulated a single crystallite orientation, again using EasySpin.

In order to mimic the effects of the effective MW irradiation strength on the off-resonance irradiation, for each DNP mechanisms, we added Gaussian broadening to the different OE, SE and CE lineshapes \cite{Shimon2012}. The effective irradiation on the SE is weakest, and therefore it was broadened with a Gaussian function with a FWHM of 0.1 MHz. The effective irradiation of the OE and CE are much larger, and therefore they were broadened with a Gaussian function with a FWHM of 8 MHz. The FWHM values were chosen in order to give the best fit for the DNP spectrum. However, from perturbation theory we know that the effective irradiation at the DQ/ZQ transition (as in the SE) is of the order of about 1/100 of the irradiation on a single quantum transition (as in the CE/OE), which is reflected in the values chosen \cite{Hovav2010,Hovav2012_CE}.

The basic DNP shapes used for the fitting were calculated by convoluting the three EPR lines with delta functions at the appropriate frequency separations. 
The OE shapes are just the EPR lines:
\begin{equation}
   \text{OE}_1(\omega_{\text{MW}})=g_1(\omega_{\text{MW}})
\end{equation}
\begin{equation}
   \text{OE}_2(\omega_{\text{MW}})=-g_2(\omega_{\text{MW}})
\end{equation}
\begin{equation}
   \text{OE}_3(\omega_{\text{MW}})=-g_3(\omega_{\text{MW}})
\end{equation}

The SE shapes are just the EPR lines convolved with delta functions at $\omega = \omega_{\text{MW}}+\omega_C$ and $\omega = \omega_{\text{MW}}-\omega_C$, giving:
\begin{equation}
    \text{SE}_1(\omega_{\text{MW}})=g_1(\omega_{\text{MW}}+\omega_C)-g_1(\omega_{\text{MW}}-\omega_C)
\end{equation}
\begin{equation}
    \text{SE}_2(\omega_{\text{MW}})=g_2(\omega_{\text{MW}}+\omega_C)-g_2(\omega_{\text{MW}}-\omega_C)
\end{equation}
\begin{equation}
    \text{SE}_3(\omega_{\text{MW}})=g_3(\omega_{\text{MW}}+\omega_C)-g_3(\omega_{\text{MW}}-\omega_C)
\end{equation}

The CE shapes are just the EPR lines convolved with delta functions at $\omega = \omega_{\text{MW}}+\omega_C$ and $\omega = \omega_{\text{MW}}-\omega_C$, and then convolved again with the two outer EPR lines, giving:
\begin{equation}
    \text{CE}_1(\omega_{\text{MW}})=g_1(\omega_{\text{MW}})*[g_1(\omega_{\text{MW}}+\omega_C)-g_1(\omega_{\text{MW}}-\omega_C)]
\end{equation}
\begin{equation}
    \text{CE}_3(\omega_{\text{MW}})=g_3(\omega_{\text{MW}})*[g_3(\omega_{\text{MW}}+\omega_C)-g_3(\omega_{\text{MW}}-\omega_C)]
\end{equation}
CE from the central EPR line was not considered because the line is not broad enough to fulfill the CE-condition of two electrons separated by the nuclear Larmor frequency.

To reproduce the experimental DNP spectrum, a linear combination of the different basic DNP shapes was calculated:

\begin{equation}
    \epsilon(\omega_{\text{MW}})=
    \epsilon_{\text{OE}}(\omega_{\text{MW}})+\epsilon_{\text{SE}}(\omega_{\text{MW}})+\epsilon_{\text{CE}}(\omega_{\text{MW}})
\end{equation}

\noindent where

\begin{eqnarray}
    \epsilon_{\text{OE}}(\omega_{\text{MW}})&=&
    \epsilon_{\text{OE,1}}(\omega_{\text{MW}})+\epsilon_{\text{OE,2}}(\omega_{\text{MW}})+ \epsilon_{\text{OE,3}}(\omega_{\text{MW}}) \nonumber \\
    &= & k_{\text{OE,1}}\text{OE}_1(\omega_{\text{MW}})+k_{\text{OE,2}}\text{OE}_2(\omega_{\text{MW}})+ \nonumber \\
    & & \hspace*{0.5in} k_{\text{OE,3}}\text{OE}_3(\omega_{\text{MW}})
\end{eqnarray}

\begin{eqnarray}
    \epsilon_{\text{SE}}(\omega_{\text{MW}})&=&
    \epsilon_{\text{SE,1}}(\omega_{\text{MW}})+\epsilon_{\text{SE,2}}(\omega_{\text{MW}})+ \epsilon_{\text{SE,3}}(\omega_{\text{MW}}) \nonumber \\
    &=& k_{\text{SE,1}}\text{SE}_1(\omega_{\text{MW}})+k_{\text{SE,2}}\text{SE}_2(\omega_{\text{MW}})+\nonumber \\
    & & \hspace*{0.5in} k_{\text{SE,3}}\text{SE}_3(\omega_{\text{MW}})
\end{eqnarray}

\begin{eqnarray}
    \epsilon_{\text{CE}}(\omega_{\text{MW}})&=& 
    \epsilon_{\text{CE,1}}(\omega_{\text{MW}})+\epsilon_{\text{CE,3}}(\omega_{\text{MW}}) \nonumber \\
    &=& k_{\text{CE,1}}\text{CE}_1(\omega_{\text{MW}})+k_{\text{CE,3}}\text{CE}_3(\omega_{\text{MW}})
\end{eqnarray}

In order to fit the DNP spectrum we manually adjusted the values of $k_{\text{OE,1}}$, $k_{\text{OE,2}}$, $k_{\text{OE,2}}$, $k_{\text{SE,1}}$, $k_{\text{SE,2}}$, $k_{\text{SE,2}}$, $k_{\text{CE,1}}$ and $k_{\text{CE,2}}$, and determined the best fit by eye. The contribution of the tCE to the fit is included in the OE terms for the two outer EPR lines, since the shapes are identical for both mechanisms.



\section{Build-up of the Thermal $^{13}$C Signal}
\label{SISec:thermal-build-up}
\noindent The buildup for the thermal signal (T$_{1}$ curve) was recorded using a saturation-recovery experiment. Because of the long times involved in this experiment, the longer times were recorded with fewer scans (64 instead of 128 or 480 scans), and therefore the intensity of these points have slightly larger error-bars. The curve was fitted with a biexponential fit, with time constants of T$_{1}^{\text{short}}= 51\pm 29$  s and T$_{1}^{\text{long}}= 481\pm 425$ s.  Note that the uncertainties in the fit are quite large due to both the low SNR in the thermal experiment and the biexponential nature of the fit.

\begin{figure}
\includegraphics[width=0.5\textwidth]{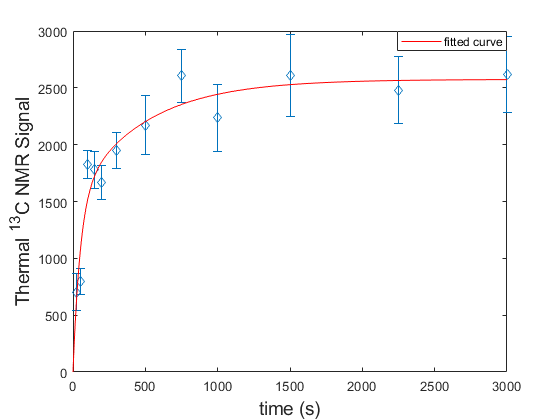}
\label{fig:Thermal-buildup}
\caption{Thermal build up of NMR signal (blue diamonds) overlaid with the biexponential fit (red line).}
\end{figure}

\section{Power Dependence of DNP}
\label{SISec:power}
\begin{figure}[ht]
\includegraphics[width=0.5\textwidth]{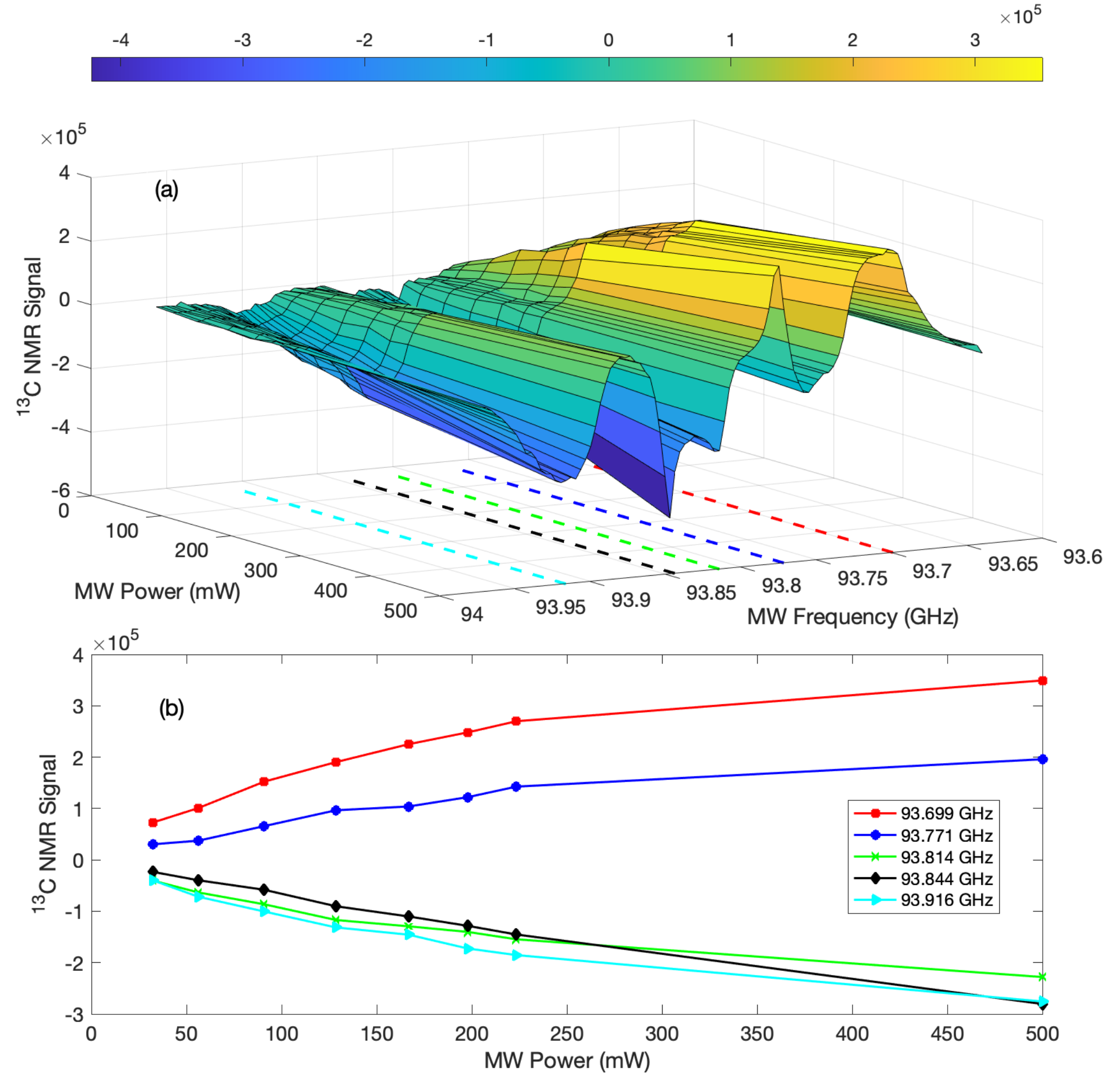}
\caption{(a) $^{13}$C NMR Signal as a function of MW power at a DNP build-up time of 300 s indicating that the DNP enhancement is still growing as a function of power at all frequencies in our experiment. (b) Power dependence at specific MW frequencies showing that the enhancement is beginning to saturate at some frequencies. \label{fig:DNP_power}}
\end{figure}

\noindent It has been reported that different DNP mechanisms have different dependence on the power of the MW irradiation. As such, we should be able to tell apart the SE, OE and CE via their power dependencies \cite{kundu2019dnp,corzilius2020high}.  Figure~\ref{fig:DNP_power} shows the DNP spectrum as a function of applied MW power at buildup times of 300 s.  For the 240 mW source, the power was varied by changing the settings of a variable attenuator.  The figure suggests that the DNP enhancement grows linearly with power at all MW irradiation frequencies up to about 240 mW suggesting that we are in a low enough power regime that the enhancement has not reached saturation for any of the DNP mechanisms \cite{siaw2016versatile}. At 500 mW the SE contributions have continued to increase linearly but the enhancement starts to saturate at other MW frequencies.

\begin{figure}
\includegraphics[width=0.5\textwidth]{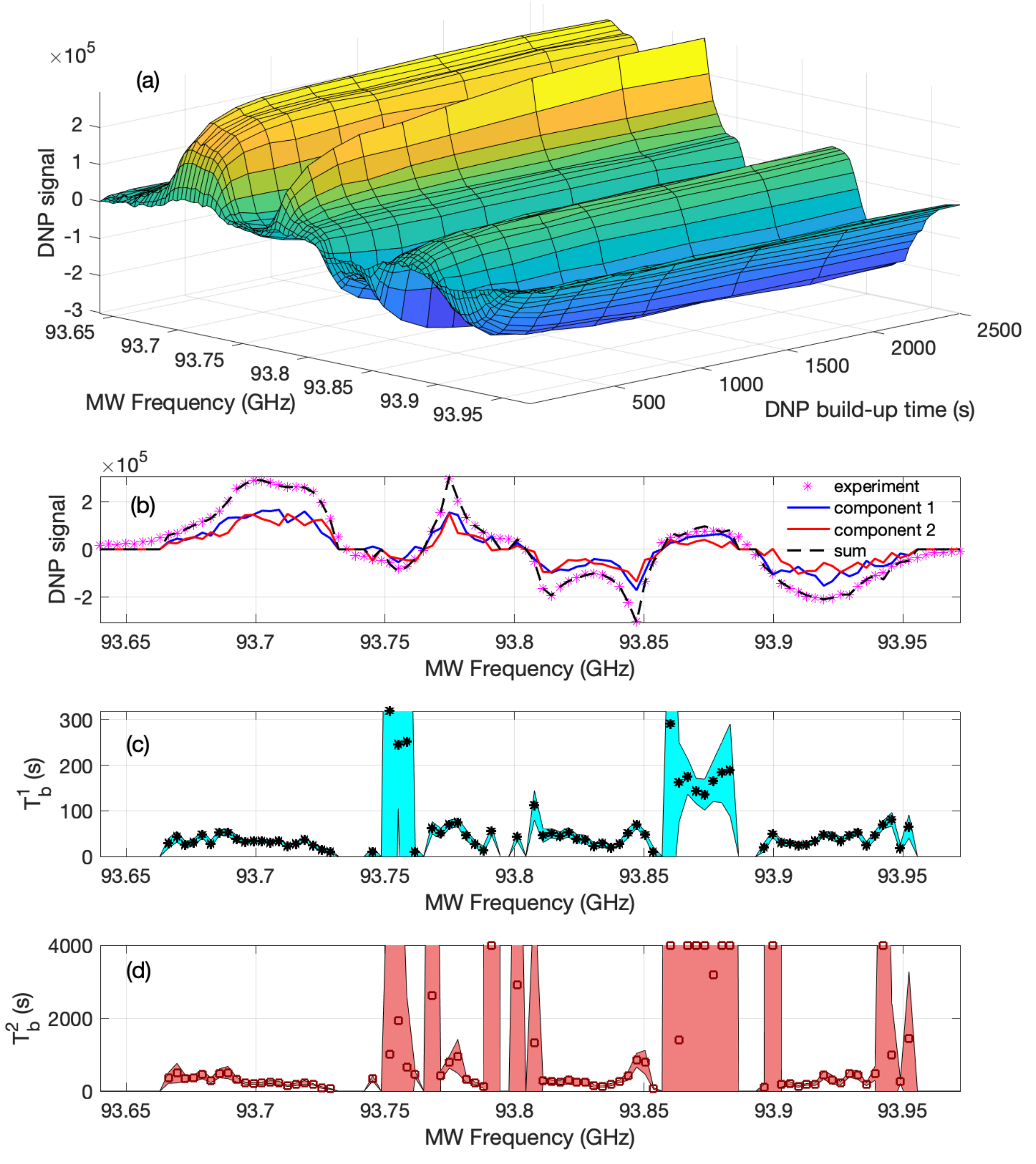}
\caption{(a) Surface Plot showing the build up of the DNP spectrum as a function of MW irradiation time.  (b) The build-up curves at each MW frequency were fit using a two-component fit.  The amplitudes of the two components and their sum are shown in (b), while (c) and (d) show the best-fit short and long time constants as a a function MW frequency.  The shaded region indicates the $\pm \sigma$ confidence interval of the fits.  The fits fail when the DNP enhancement is low and where the enhancement switches sign. There is significant uncertainty associated with some of the longer time constants since data was only acquired up to a maximum build-up time of 3000 s.}
\label{fig:DNP-buildup}
\end{figure}

\section{DNP Buildup Times}
\label{SISec:DNP-build-up}
\noindent In addition, to compare the various DNP mechanisms, we also characterized the build-up time of the hyperpolarization as a function of MW frequency at 240 mW up to 3000 s as shown in Figure \ref{fig:DNP-buildup}.  
Almost all the buildup curves could be fit with a two-component fit, T$_{bu}^{\text{short}}$ and T$_{bu}^{\text{long}}$. This biexponential behavior suggests that the two components consist of $^{13}$C spins directly hyperfine-coupled to an adjacent P1 center (short time constant) and more distant spins that rely on nuclear spin diffusion to mediate both T$_1$ relaxation and DNP buildup (long time constant) \cite{Ramanathan-2008}.  Only a single exponential fit was required at a few MW excitation frequencies.  The fits fail when the DNP enhancement is low such as on the edges of the DNP spectrum and in locations where the enhancement switches sign. The amplitudes and time constants of the two components are also shown in the figure. The shaded region indicates the $\pm \sigma$ confidence interval of the fitted time constants. There is significant uncertainty associated with some of the longer time constant since data was only acquired up to a maximum build-up time of 3000 s.

At most MW excitation frequencies, the short component is on the order of 50 s and the long component on the order of 500 s -- similar to the thermal build-up times. At most MW frequencies, the amplitudes of the short and long time constant components are nearly the same, except at the outer lobes around 93.7 GHz and 93.92 GHz, where the short time component is seen to be significantly larger than the long-time component.  The buildup was observed to be particularly slow for the SE peaks.

The fact that T$_{1}^{\text{short}}\approx$ T$_{bu}^{\text{short}}$ and T$_1^{\text{long}}\approx$ T$_{bu}^{\text{long}}$ is another indication that we do not have enough MW power to drive the DNP at a rate that is faster than T$_{1}$ \cite{Hovav2010,kundu2019dnp}.

\section{Simulation of DNP Mechanisms} \label{sec:simulations}
\noindent We performed numerical simulations of the spin dynamics of two small spin systems, an {e-$^{14}$N} system and an {e-$^{14}$N-$^{13}$C} system. The simulations were based on the method described by the Vega group \cite{Hovav2012,Hovav2012_CE,Hovav2014,Shimon2012}. Numerical simulations involving e-$^{14}$N and e-$^{14}$N-$^{13}$C systems are described in Shimon et al.\ in the context of the role the $^{14}$N nucleus plays in DNP enhancement of spin half nuclei using Nitroxide radicals \cite{Shimon2012}.

\subsection{Simulation of the e-$^{14}$N System}\label{SIsec:eNsystem}

\noindent We begin by looking at an {e-$^{14}$N} system, containing one electron that is strongly hyperfine coupled to a $^{14}$N nucleus. In the MW rotating frame, the Hamiltonian of the system without the MW irradiation is given by:
\[H_{0}^N=\Delta\omega _eS_z-\omega _NI_z^N+A_{z}^NS_zI_z^N+A_{x}^NS_zI_x^N+A_{y}^NS_zI_y^N\]
where $S$ and $I^N$ are the electron and nitrogen spins, respectively. $\omega_e=\omega_e-\omega_{\text{MW}}$ is the electron off-resonance (which is determined by the g-tensor), $\omega_N$ is the $^{14}$N Larmor frequency. $A_z^N$ is the secular part of the e-N hyperfine interaction (which includes both isotropic and anisotropic shifts) and A$_{x}^N$ and A$_{y}^N$ are the pseudo-secular terms of the dipolar hyperfine interaction (which is anisotropic). Here, we simulated each P1 center as an e-$^{14}$N system, with an isotropic g=2.0024 and hyperfine coupling strengths with the $^{14}$N nucleus with principal axis components $A_{xx}^N = A_{yy}^N = 82$ MHz, $A_{zz}^N$ = 114 MHz \cite{Loubser-1978}. A single crystal orientation was chosen, such that $A_z^N$ = 98 MHz, and  $A_x^N$ = -16 MHz and $A_y^N$ = 0 MHz. In the rotating frame, the MW irradiation term is

\begin{equation}
   \omega_{\text{MW}}=\omega _1S_z.
\end{equation}

\noindent Note, that the nuclear quadrupolar interaction is not included in these simulations.  First, we diagonalize $H_0^N$ according to

\begin{equation}
    H_{0}^{Nd}=V^{-1}H_{0}^NV
\end{equation}

The $A_z^N$ term splits the EPR transition $|\uparrow_e\rangle \leftrightarrow |\downarrow_e\rangle$ into three transitions, according the nitrogen spin eigenstates $|\chi_N\rangle=|-1\rangle$, $|0\rangle$ or $|1\rangle $, such that the electron transitions can be written as $|\uparrow_e,\chi_N\rangle\leftrightarrow|\downarrow_e,\chi_N\rangle$, separated by $A_z^N$.
The $A_x^N$ and $A_y^N$ terms result in weak state mixing of adjacent (i.e., $\Delta m_I=\pm1$) nitrogen spin states within each electron manifold, such that:
\begin{eqnarray}
   |\chi_e, \pm \tilde{1}\rangle & = & c_N|\chi_e,\pm1\rangle+s_N|\chi_e,0\rangle \\
   |\chi_e,\tilde{0}\rangle & = & c_N|\chi_e,0\rangle+s_N|\chi_e,+1\rangle-s_N|\chi_e,-1\rangle 
\end{eqnarray}
\noindent where $|\chi_e\rangle=|\uparrow_e\rangle$ or $|\downarrow_e\rangle$,  $c_N=\cos\zeta_N$ and $s_N=\sin\zeta_N$, $\tan2\zeta_N=\sqrt{(A_x^N)^2+(A_y^N)^2}/[8(\omega_N+A_z^N)]$, if $\sqrt{(A_x^N)^2+(A_y^N)^2} \ll \omega_N$ and $A_z^N$, from perturbation theory \cite{Schweiger2001}.

Next, we transform the MW irradiation term to the same frame, according to

\begin{equation}
    H^d_{\text{MW}}=V^{-1}H_{\text{MW}}V
\end{equation}

The weak state mixing described above also results in weak effective irradiation between double quantum and zero quantum transitions, which give the $^{14}$N-SE-DNP enhancement. The strength of the effective irradiation is proportional to the state-mixing, such that it is $\omega _{1,\text{eff}}^N=|s_N|\omega_{1}$ for the $^{14}$N-SE-DNP and even weaker for second order $^{14}$N-SE-DNP effects.

\begin{figure}
\includegraphics[width=0.48\textwidth]{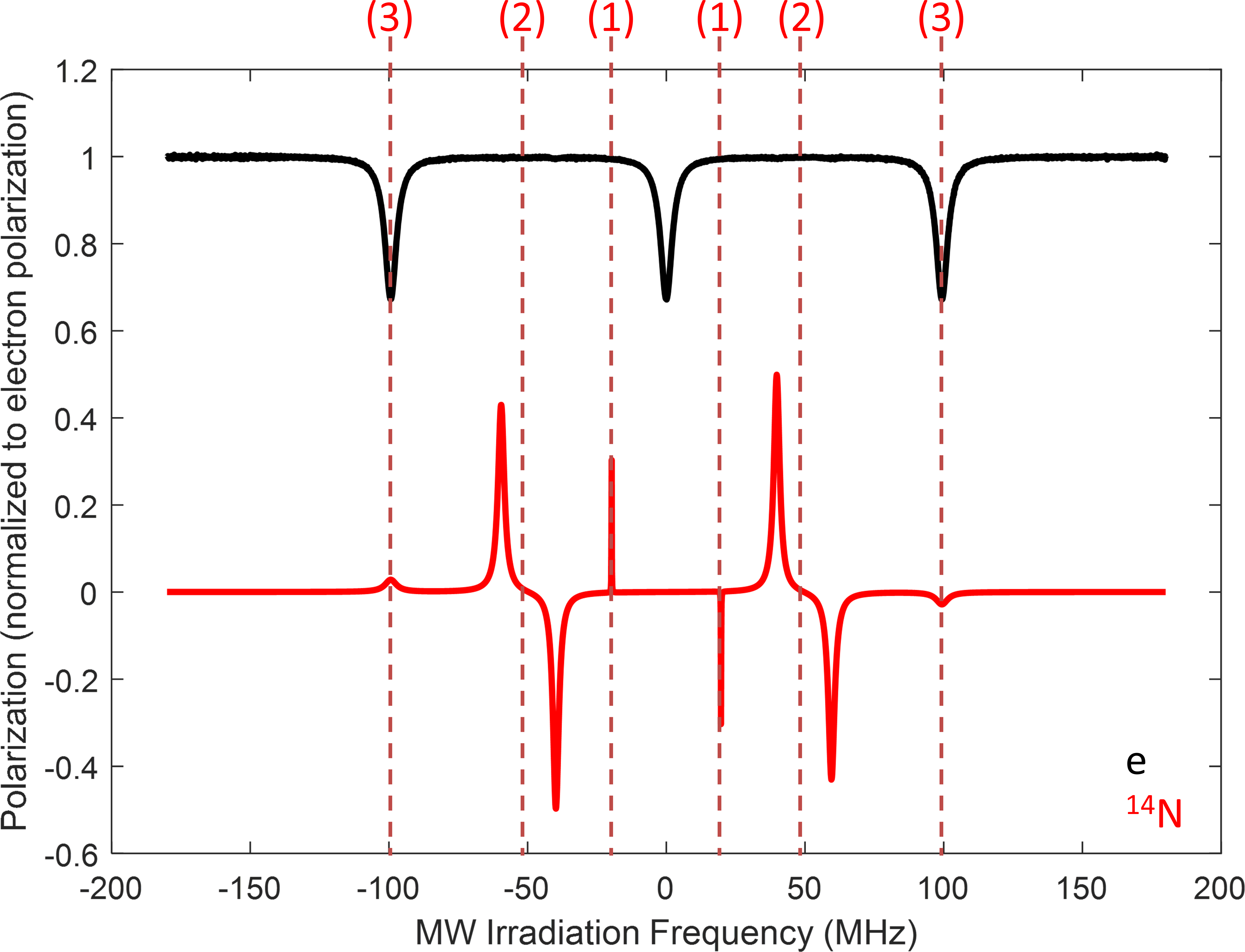}
\caption{Simulated electron (black) and $^{14}$N (red) polarizations as a function of the MW irradiation frequency for an e-$^{14}$N spin system. The polarizations are  normalized according to the steady state electron polarization. The MW irradiation frequency is plotted in MHz for convenience, and is referenced to the electron Larmor frequency, $\omega_e=$94 GHz. The simulation parameters: $\omega_e=$94 GHz, $\omega_N=$10 MHz, $A_z^N=$98 MHz, $A_z^N=$98 MHz, $A_x^N$=-16 MHz, $A_y^N=$0 MHz, $\omega_1=$0.5 MHz, T$_{1e}=$10$^3$ ms, T$_{1N}=$1 s, T$_{2e}=$10 $\mu$s and T$_{2N}=$100 $\mu$s. No cross relaxation was added to the system.
\label{fig:DNP_spectrum_sim1}}
\end{figure}

In Fig.\ \ref{fig:DNP_spectrum_sim1} we plot the steady state $^{14}$N nuclear polarization, $\text{P}_N(\omega_{\text{MW}})$, and the electron polarization, $\text{P}_e(\omega_{\text{MW}})$, during MW irradiation as a function of the frequency of the MW irradiation. Here we normalize all electron and nuclear polarizations to the value of the electron polarization at thermal equilibrium, $\text{P}_e^{\text{eq}}=1$ (i.e., when no MW irradiation is applied). Concentrating on the electron, it is clear that when the MW irradiation frequency is far off resonance the electron polarization is at its largest value. When the MW irradiation is on-resonance with one of the electron transitions the electron polarization is reduced (i.e., the electron is partially saturated). Far off resonance $\text{P}_N(\omega_{\text{MW}})=\text{P}_N^{\text{eq}}\approx 0$ because the nuclear polarization is negligible compared to the electron polarization ($\text{P}_e^{\text{eq}}\gg \text{P}_N^{\text{eq}}$). At certain frequencies the $\text{P}_N(\omega_{\text{MW}})\neq 0$ is enhanced due to the small state mixing between the nitrogen spin states, which is a result of the $A_z^N$ term of the hyperfine interaction.

Concentrating on the $^{14}$N-DNP spectrum in Fig.\ \ref{fig:DNP_spectrum_sim1}, it can be clearly seen that there are three distinct DNP features (marked 1,2 and 3 in the figure):

\begin{enumerate}
    \item Two narrow features appearing at $\omega \approx \omega_e \pm 2\omega_N$ are second order SE-DNP transitions. Positive enhancement is achieved when irradiating on the $|\alpha_e, \tilde{1}\rangle \leftrightarrow|\beta_e, -\tilde{1}\rangle$ transition, and negative enhancement is achieved when irradiating between $|\alpha_e, -\tilde{1}\rangle \leftrightarrow|\beta_e, \tilde{1}\rangle$.
    
    \item Four broader features appearing at $\omega = \omega_e \pm \omega_N \pm A_z^N/2$ and $\omega = \omega_e \pm \omega_N \mp A_z^N/2$ are the SE transitions of the central EPR line, where the separation between positive and negative enhancement is $2\omega_N + A_z^N$ or $2\omega_N - A_z^N$. The outer SE pair is formed as a result of irradiation on the $|\alpha_e, \tilde{0}\rangle \leftrightarrow|\beta_e, -\tilde{1}\rangle$ and $|\alpha_e, \tilde{0}\rangle \leftrightarrow|\beta_e, +\tilde{1}\rangle$ transitions. The inner SE pair is formed as a result of irradiation on the $|\alpha_e, -\tilde{1}\rangle \leftrightarrow|\beta_e, \tilde{0}\rangle$ and the $|\alpha_e, \tilde{1}\rangle \leftrightarrow|\beta_e, \tilde{0}\rangle$. The two SE pairs are centered around $\omega_e$, but they have opposite signs, such that signs of the enhancement of the inner SE pair are opposite of those of the outer SE pair.
    
    \item The two broader features at $\omega = \omega_e \pm A_z^N$ appear on resonance on the electron, as one would expect from the Overhauser effect, but are in fact due to the SE. In this case, the state-mixing described in Equations 2-3 in the main text results in electron-nitrogen cross-relaxation terms that connect the nitrogen states within each electron manifold. When irradiating on the low frequency electron line (i.e. on the $|\alpha_e, -\tilde{1}\rangle \leftrightarrow|\beta_e, -\tilde{1}\rangle$ transition) and then positive enhancement is achieved. In a similar manner, when irradiating on the high frequency electron line (i.e. on the $|\alpha_e, +\tilde{1}\rangle \leftrightarrow|\beta_e, +\tilde{1}\rangle$ transition), $T_{1ZQ}$ forms on the $|\alpha_e, +\tilde{1}\rangle \leftrightarrow|\beta_e, \tilde{0}\rangle$ transition and then negative enhancement is achieved. At the center of the DNP spectrum (when irradiating directly at $\omega$, no enhancement is achieved).
\end{enumerate}

\subsection{Simulation of the e-$^{14}$N-$^{13}$C system}


\noindent Next, we add a $^{13}$C spin to the system described above. The Hamiltonian of this system is given by:

\[H_0^{NC}=H_{0}^N-\omega _CI_z^C+A_{x}^CS_zI_x^C+A_{y}^CS_zI_y^C\]
where $I^C$ is the carbon spin operator, $\omega_C$ is the $^{13}$C Larmor frequency and $A_{x}^C$ and $A_{y}^C$ are the pseudo-secular terms of the dipolar hyperfine interaction. The other terms were defined above. Note that no e-C secular hyperfine or N-C dipolar interactions were added to the simulation. Here, $A_{x}^C=A_{y}^C$.

Again we diagonalize the Hamiltonian $H_{0}^{NC}$, and then transform $H_{\text{MW}}$ to the same frame, as described above. The $A_{x}^C=A_{y}^C$ term results in weak state mixing of carbon spin spin states within each electron-nitrogen manifold, such that:

\begin{eqnarray}
   |\chi_e,\tilde{\chi}_N,\alpha _C^*\rangle & = & c_C|\chi_e,\tilde{\chi}_N,\alpha _C\rangle +s_C|\chi_e,\tilde{\chi}_N,\beta _C\rangle \label{eq:mixing3}\\
   |\chi_e,\tilde{\chi}_N,\beta _C^*\rangle & =c &_C|\chi_e,\tilde{\chi}_N,\beta _C\rangle-s_C|\chi_e,\tilde{\chi}_N,\alpha _C\rangle  \label{eq:mixing4}
\end{eqnarray}

\noindent where $\chi_e=|\alpha_e\rangle$ or $|\beta_e\rangle $, $\tilde{\chi}_N=|\pm\tilde{1}\rangle$ or $\tilde{\chi}_N=|\tilde{0}\rangle$ and $c_C=\cos\zeta_C$ and $s_C=\sin\zeta_C$ and $\tan2\zeta_N=\sqrt{(A_x^C)^2+(A_x^C)^2}/[8\omega_C]$, if $\sqrt{(A_x^C)^2+(A_x^C)^2} \ll \omega_N$, from perturbation theory \cite{Schweiger2001}.

\begin{figure}
\includegraphics[width=0.48\textwidth]{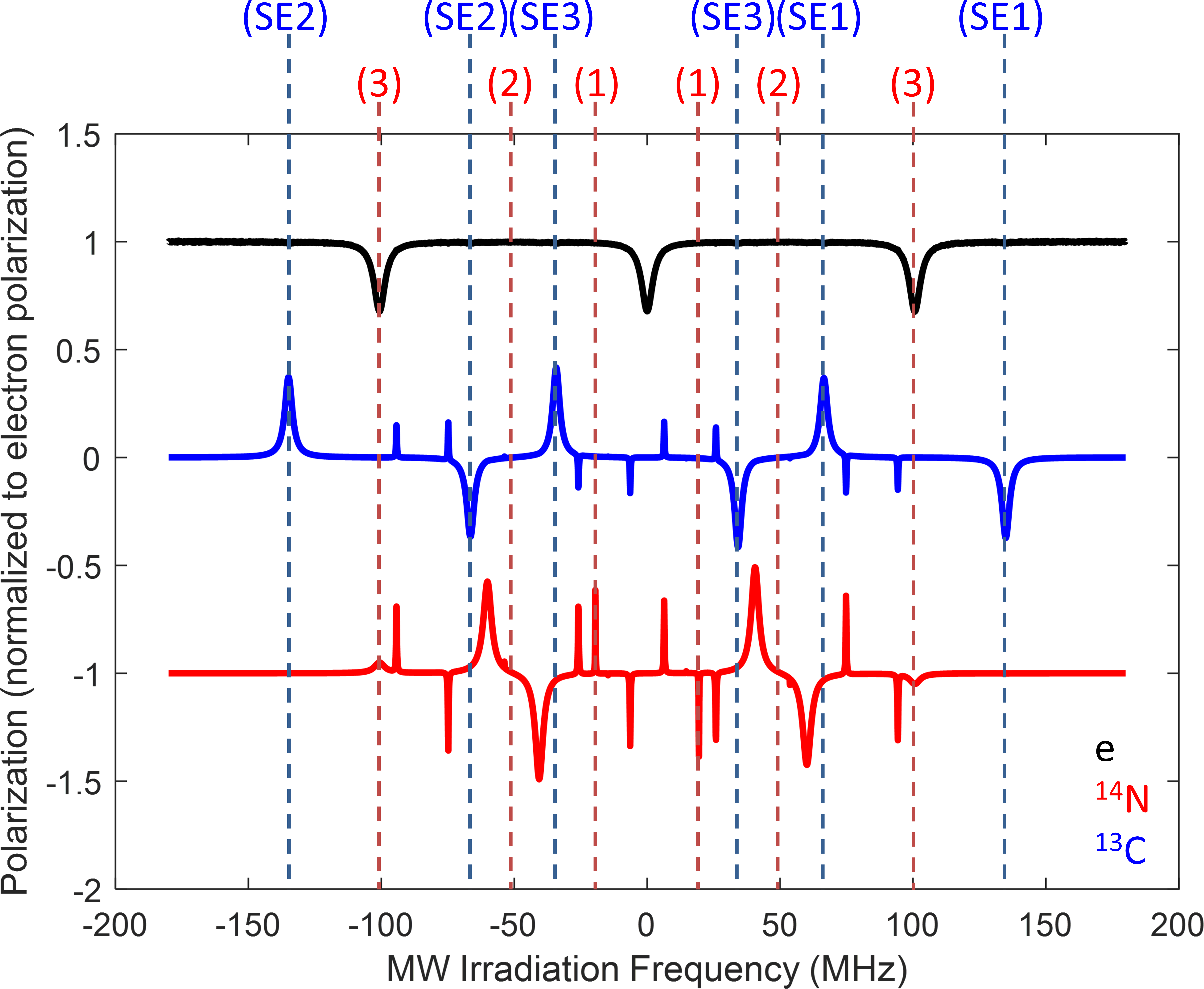}
\caption{Simulated electron (black), $^{14}$N (red) and $^{13}$C polarizations as a function of the MW irradiation frequency for an e-$^{14}$N-$^{13}$C spin system. The polarizations are  normalized according to the steady state electron polarization. The $^{14}$N DNP spectrum was shifted in the y-axis for clarity. The MW irradiation frequency plotted in MHz for convenience, and is referenced to the electron Larmor frequency, $\omega_e=$94 GHz. The simulation parameters: $\Delta\omega_e=$94 GHz, $\omega_N=$10 MHz, $\omega_C=$34 MHz, $A_z^N=$98 MHz, $A_x^N$=-16 MHz, $A_y^N=$0 MHz, $A_x^C=A_y^C=$5.4 MHz (corresponds to a 15.4 nm distance), $\omega_1=$0.5 MHz, T$_{1e}=$10$^3$ ms, T$_{1N}=$1 s, T$_{1C}=$100 s, T$_{2e}=$10 $\mu$s and T$_{2N}=$T$_{2C}=$100 $\mu$s. No cross relaxation was added to the system. \label{fig:DNP_spectrum_sim2}}
\end{figure}

The weak carbon state-mixing described above also results in weak effective-irradiation on the e-C DQ and ZQ transitions, which results in $^{13}$C-SE-DNP enhancement, as can be seen in Fig.\ \ref{fig:DNP_spectrum_sim2}. The strength of the effective irradiation on the $^{13}$C-SE transitions is proportional to the state-mixing, such that it is $\omega _{1,\text{eff}}^{C}=|s_C|\omega _{1}$ for the $^{13}$C-SE-DNP and even weaker for second order $^{13}$C-SE-DNP effects.

As can be seen in Fig.\ \ref{fig:DNP_spectrum_sim2}, the addition of the $^{13}$C nucleus only slightly alters the $^{14}$N-DNP spectrum. The same three features described above are still visible here (marked 1,2 and 3 in the figure), in addition to several new and narrow DNP features. The additional narrow features are a result of second-order SE DNP lines involving the electron, the $^{14}$N and the $^{13}$C nuclei. For example, MW irradiation on the transition $|\alpha_e,\tilde{1},\alpha _C^*\rangle\leftrightarrow|\beta_e,\tilde{0},\beta _C^*\rangle$ will result in DNP enhancement of both the nitrogen and the carbon. As such, both the $^{14}$N and the $^{13}$C nuclei show DNP enhancement at these SE mechanisms. These effects are beyond the scope of the current work. 

In addition to the narrow second-order SE DNP lines, the $^{13}$C-DNP spectrum exhibits three $^{13}$C-SE pairs, one from each electron line (marked SE1, SE2 and SE3 in the figure). These SEs appear at $\omega = \Delta\omega_e + A_z^N \pm \omega_C$ (SE1), $\omega = \Delta\omega_e - A_z^N \pm \omega_C$ (SE2) and $\omega = \Delta\omega_e \pm \omega_C$ (SE3). The $^{13}$C-SE-DNP mechanisms result in positive enhancement when irradiating on the $|\beta _e,\tilde{\chi}_N,\alpha _C^*\rangle \leftrightarrow|\alpha_e,\tilde{\chi}_N,\beta _C^*\rangle$ transition and negative enhancement when irradiating on the $|\beta _e,\tilde{\chi}_N,\beta _C^*\rangle \leftrightarrow|\alpha_e,\tilde{\chi}_N,\alpha _C^*\rangle$ transition, for $\tilde{\chi}_N=-\tilde{1}$, $\tilde{0}$ or $\tilde{1}$.

In summary, the DNP simulations for an e-$^{14}$N-$^{13}$C system exhibit a $^{13}$C-SE for each of the three e-$^{14}$N manifolds. In each case, the $^{13}$C-SE is independent of the $^{14}$N spin state, and does not involve the $^{14}$N spin. Using this observation, we justify the convolution method of simulating the $^{13}$C-DNP mechanisms (OE, SE and CE) for each e-$^{14}$N manifold separately (described above). Higher order $^{13}$C-DNP enhancement features that involve both the $^{14}$N spin as well as the $^{13}$C spin are much more sensitive to the MW irradiation strength, and therefore, we are able to neglect them in our analysis of the experimental results.

\section{Low-Field Pulse-EPR Spectrometer and Data Processing}
\label{SISec:PeprDetails}

The resonance frequency of 2.5 GHz is achieved by mixing a 2.476 GHz local oscillator with the output from a Tektronix AWG 7052 which is used to define pulse waveforms on a 24 MHz carrier. The pulses are amplified to approximately +27 dBm, then transmitted through a circulator to an antenna. At the end of the coax line is an exposed wire which capacitively couples to a shielded loop-gap resonator \cite{froncisz_loop-gap_1982} (inspired by a previously published design \cite{Joshi-2020}) that is 1.27 cm long and has an inner diameter of 0.73 cm. The received echo signal is amplified, mixed with the local oscillator and sampled at 400 MHz with a digitizer. In-phase and quadrature components of the pulses and echoes are transmitted and received, respectively, via IQ mixers. On-board averaging is performed on the digitizer, which returns to the lab computer the averaged echo waveform for a given set of experiment parameters. For each set of experiment parameters, the phase of the pulses is cycled through 0, $\pi/2$, $\pi$, and $3\pi/2$, and the resulting waveforms are phase shifted and added constructively in post-processing, thereby averaging out phase-independent systematic noise. The echo is then multiplied by a 1.0 $\mu$s Blackman window to suppress noise outside of the echo in the time-domain buffer. The echo is then demodulated to baseband and Fourier transformed. The echo amplitude is then obtained by integrating over the Fourier transform of the echo with an integration range of 4.0 MHz. The in-phase and quadrature components of the integrated signal are the real and imaginary components of the echo amplitude. The uncertainties are obtained by calculating the standard deviation of the Fourier transformed buffer in the 2 MHz ranges above and below the 4 MHz range used to obtain the integrated signal, then multiplying by the square root of the number of points in the signal integration part of the buffer.
In the Hahn-echo experiments, the values of $\tau$ were chosen such that values of $\log_{10}(\tau)$ had a uniform random distribution, and the order in which they were performed was random as well. For the diamond powder, $\tau$ ranged from 1.95 $\mu$s to 98.1 $\mu$s. The lengths of the $\pi/2$ and $\pi$ pulses were 200 ns and 400 ns respectively. For the single crystal macle-cut diamond, two data sets were combined: one with 28 points with $\tau$ ranging from 1.445 $\mu$s to 99.8065 $\mu$s and one with 37 points with $\tau$ ranging from 1.445 $\mu$s to 999.63 $\mu$s. The lengths of the $\pi/2$ and $\pi$ pulses were 250 ns and 500 ns respectively. The delay between the end of one trial and the beginning of the repeat of that trial was 10 ms. All echo magnitudes were normalized by dividing by the number of averages and then dividing by the maximum strength echo obtained with the powder. For each set of parameters, $2^{14}$ trials were averaged on-board the digitizer, and with the phase-cycling described above, the total number of averages is $2^{16}$.
The inversion recovery experiments used similar randomization to that of the Hahn-echo experiments. The values of $\tau_1$ were chosen such that values of $\log_{10}(\tau_1)$ had a uniform random distribution and random order. For each set of parameters tested, back-to-back tests are performed with the inverting pulse off (amplitude set to 0) and then on, giving echo amplitudes $S_\text{off}$ and $S_\text{on}$, respectively. The same phase correction is applied to both by setting $S_\text{off}$ to be purely real. Then the inversion recovery signal plotted in Figure \ref{fig:EPR_T1T2_fit}(b) is $S_\text{IR} = \textrm{Re}[S_\text{on}]/\textrm{Re}[S_\text{off}]$. For the diamond powder and the single crystal, $\tau_1$ ranged from 1 $\mu$s to 60 ms, and $\tau_2$ was set such that the time between the middles of pulses 2 and 3 was 3 $\mu$s. The delay between the end of one trial and the beginning of the repeat of that trial was 100 ms. For the powder, the pulses were the same as described in the above paragraph, but for the single crystal, a recalibration was performed which led to $\pi/2$ pulses of 330 ns and $\pi$ pulses of 660 ns being used. The need for longer pulse durations for the single crystal is due to a different resonator and sample holder being used.

\end{document}